\documentclass[12pt,preprint]{aastex}

\newcommand{\age}{\stackrel{>}{\sim}}
\newcommand{\ale}{\stackrel{<}{\sim}}

\def\siml{\hspace{1ex} ^{<} \hspace{-2.5mm}_{\sim} \hspace{1ex}}

\newcommand{\eq}{\begin{equation}}

\def\gapp{\ \lower 3pt\hbox{${\buildrel > \over \sim}$}\ }
\def\lapp{\ \lower 3pt\hbox{${\buildrel < \over \sim}$}\ }
\def\ba{{\bf a}}
\def\bb{{\bf b}}
\def\bp{{\bf p}}
\def\mybv{{\bf v}}
\def\bA{{\bf A}}
\def\bB{{\bf B}}
\def\bJ{{\bf J}}
\def\bahat{\hat{\bf a}}
\def\bAhat{\hat{\bf A}}
\def\bbhat{\hat{\bf b}}

\def\bphat{\hat{\bf p}}
\def\bBhat{\hat{\bf B}}
\def\ahat{\hat{a}}

\def\bhat{\hat{b}}

\def\rp{r_{\rm p}}
\def\modD{\vert D\vert}
\def\modDelta{\vert\Delta\vert}

\newcommand{\ben}{\begin{enumerate}}
\newcommand{\een}{\end{enumerate}}

\newcommand{\be}{\begin{equation}}
\newcommand{\ee}{\end{equation}}
\newcommand{\bea}{\begin{eqnarray}}
\newcommand{\eea}{\end{eqnarray}}
\newcommand{\bean}{\begin{eqnarray*}}
\newcommand{\eean}{\end{eqnarray*}}

\newcommand{\br}{{\bf r}}

\shorttitle{Dynamics of Planetary Systems in Star Clusters}
\shortauthors{Spurzem et al.}

\begin{document}

\title{Dynamics of Planetary Systems in Star Clusters}

\author{R. Spurzem$^{1}$, M. Giersz$^2$, D.C. Heggie$^3$,
D. N. C. Lin$^{4,5}$}

\affil{1) Astronomisches Rechen-Institut, Zentrum f\"ur Astronomie,
University of Heidelberg,
M\"onchhofstr. 12-14, 69120 Heidelberg, Germany \\
2) Nicolaus Copernicus Astronomical Centre, Bartycka 18, 
00-716 Warszawa, Poland \\
3) University of Edinburgh, School of Mathematics and Maxwell
Institute for Mathematical Sciences, King's
Buildings, Mayfield Rd., Edinburgh EH9~3JZ, Scotland, UK \\
4) UCO/Lick Observatory, University of California, Santa Cruz, CA
95064, USA\\
5) Kavli Institute for Astronomy and Astrophysics, Peking University, 
Beijing, China}

\email{spurzem@ari.uni-heidelberg.de}


\begin{abstract}
At least 10-15\% of nearby sun-like stars have known Jupiter-mass
planets.  In contrast, very few planets are found in mature open and
globular clusters such as the Hyades and 47 Tuc. We explore here the
possibility that this dichotomy is due to the post-formation disruption
of planetary systems associated with the stellar encounters in
long-lived clusters.  One supporting piece of evidence for this scenario is the
discovery of freely floating low-mass objects in star forming
regions. We use two independent numerical approaches, a hybrid Monte
Carlo and a direct $N$-body method, to simulate the impact of the
encounters.  We show that the results of numerical simulations are in
reasonable agreement with analytical determinations in the adiabatic
and impulsive limits. They indicate that distant stellar encounters
generally do not significantly modify the compact and nearly circular
orbits.  However, moderately close stellar encounters, which are
likely to occur in dense clusters, can excite planets' orbital
eccentricity and induce dynamical instability in systems which are
closely packed with multiple planets.  The disruption of planetary
systems occurs primarily through occasional nearly parabolic, non adiabatic
encounters, though eccentricity of the planets evolves through repeated
hyperbolic adiabatic encounters which accumulate small-amplitude
changes. The detached planets are generally retained by the potential
of their host clusters as free floaters in young stellar clusters such
as $\sigma$ Orionis. We compute effective cross sections for the
dissolution of planetary systems and show that, for all initial
eccentricities, dissolution occurs on time scales which are longer than the
dispersion of small stellar associations, but shorter than the age of
typical open and globular clusters.  Although it is much more difficult to disrupt
short-period planets, close encounters can excite modest
eccentricity among them, such that subsequent tidal dissipation leads to
orbital decay, tidal inflation, and even disruption of the close-in
planets.
\end{abstract}


\section{Introduction}
The most successful observational method for planet searches has been
the method of radial velocity surveys. Among over 1000 target stars, Jupiter-
mass planets, with period up to a few years, have been found around
more than 10 \% of them (Cumming {\it et al.} 2007). This fraction is
likely to increase with follow-up high-precision measurements of
trends in the radial velocity curves which reveal suggestive signals
for many additional planetary companions.

This prolific success in the discovery of planets around these field
stars is in sharp contrast to the rarity of planets around stars in
both open and globular clusters. Despite several attempts to search
for planets in the Hyades (Guenther {\it et al.} 2005), only one
planet is found around one star (Sato {\it et al.} 2007). It is
tempting to attribute this dichotomy to an environmental influence on
the proficiency of planet formation. For example, Bonnell et
al. (2001) suggested that, if the cluster went through an initial high
density phase, close stellar encounters during the planetary formation
epoch may have tidally truncated the disk and eliminated the domain of
giant planet formation. This argument, however, is weakened by the
poorly known time scale and location of giant planet formation. It
also does not take into account the possibility of subsequent stellar
accretion of the tidally stripped gas which may provide a protracted
environment for planet formation. The discovery of planets in several
binary systems with separation of a few AU's (Queloz {\it et al.}
2000, Hatzes {\it et al.} 2003, Zucker {\it et al.} 2004, Correia {\it
et al.} 2007) is another sign that perturbations which are induced by
neighboring stars alone cannot be an efficient mechanism to quench
planet formation.

The current paradigm for star formation theory is based on the
realization that field stars are generally formed in dense gaseous
environments which are similar to, but slightly less massive and
concentrated than the progenitor of the longer-lived open
clusters. In many dense star-forming molecular clouds, the fraction of
stars with protostellar disks depends more sensitively on the stellar
age rather than the dynamical properties of their host clusters
(Hillenbrand 2005). Even in dense regions which contain massive stars,
such as the Orion nebula, the silhouettes of protostellar disks are
commonly found (O'Dell {\it et al.} 1993). Traces of radioactive decay
products in the most primitive chondritic meteorites provide strong
evidence that the formation of the solar system was preceeded by a
nearby supernova event (Cameron \& Truran 1977). Since the massive
progenitors of supernovae are only found in relatively dense molecular
clouds, it is likely that the solar system actually formed in a
stellar aggregate with $\sim 2000$ members (Adams \& Laughlin
2001). These circumstantial pieces of evidence support the conjecture that
planets may not form less prolifically in open clusters than the
progenitors of the present-day field stars.

An alternative scenario for the rarity of planet detection in open
clusters may be attributed to post-formation dynamics. There is
supporting circumstantial evidence for dynamical disruption of
planetary systems during the infancy of their host stars in dense
molecular clouds. A population of ``freely floating'' planets, with
5-15 Jupiter masses, has been found in the young cluster $\sigma$
Orionis (Zapatero-Osorio et al. 2000, Lucas et al.  2001). If these
free floaters indeed formed near some host stars, a reliable census of
their frequency in a wide range of star forming regions would provide
a calibration on the critical conditions for planetary system
disruption. However, this task is particularly challenging because
their detection and distinction from bound planets by microlensing
observations are generally difficult (c.f. Han 2006).

In contrast to open clusters,  stellar associations which
provided the birth place for the known planet-bearing field stars
probably dispersed under the combined action of internal dynamical
relaxation, stellar evolution, and the tidal perturbation of their
host galaxies over a time scale $\sim 10^8$ yr. During this epoch,
the planetary-retention probability of individual stars is determined by a
competition between cluster disruption and planetary-system break
up. The planetary-system dichotomy would be explained if stellar
interactions and encounters affect the stability and dynamical
evolution of their companion planetary systems on a time scale longer
than the associations' dispersal time scale but shorter than the age
of typical open clusters.

On the theoretical side, Smith \& Bonnell (2001) inferred that only a
small fraction of planets would become detached from their host stars
during the characteristic life span of young clusters and
association. Their results are obtained using the equations of
restricted three body motion to approximate encounters between
planetary systems with nearly circular orbits and single stars. They
claimed that the few detached planets would escape from open and young
clusters because the planets' recoil speed is generally large compared
with the clusters' velocity dispersion. These conclusions are modified
if the perturbers are binary stars. This problem has been
systematically addressed with a Monte Carlo approach by Laughlin \&
Adams (1998) in which they focused their attention on the effect of
four body scattering (a binary star, and a planet). They averaged the
stellar density over the cluster to obtain effective cross sections
for the disruption of a Jupiter-like planetary orbit by all the binary
stars. Also Davies \& Sigurdsson (2001) included in their Monte Carlo
study encounters with binaries.

In these early studies, a restricted range of planetary orbits was
under consideration.  For example, Laughlin and Adams focused their
attention on planets which were initially on a 5 AU circular orbit
around the host star.  However, the known extra solar planets
have a semi-major axis ranging from 0.05 AU to several AU. Those with
periods longer than a week have a nearly uniform eccentricity
distribution up to unity (cf. Cumming {\it et al.}, in prep.).  Although
planets are probably formed in their nascent disks beyond the snow
line with nearly circular orbits (Ida \& Lin 2004), their eccentricity
may be excited shortly after their formation through their interaction
with their nascent disks (Goldreich \& Sari 2003, Creswell {\it et
al.} 2007) or with each other (Rasio \& Ford 1996, Lin \& Ida
1997). Dynamical instability and relaxation is particularly effective
in exciting the planets' eccentricity especially in closely-packed
multiple-planet systems (Papaloizou \& Terquem 2002, Zhou {\it et al}
2007, Juric \& Tremaine 2007, Ford \& Rasio 2007).

For most non-disruptive encounters, the stellar perturbation
nonetheless modifies the eccentricities of the planets.
Individual planets with modestly eccentric orbits may be more
vulnerable to disruption by subsequent stellar perturbations. Repeated stellar interactions may be
particularly damaging to the survival of planetary systems.
Observational data also indicate that a significant fraction {of
  known planets}, if not
most of them,  have additional planetary siblings around their
host stars (Fischer et al. 2001). Although the present-day separation
in these known systems is generally wide and they are intrinsically
stable, stellar perturbation in the past can provide an additional avenue
for inducing their dynamical relaxation. The cumulative
excitation of eccentricity, which results from both close and distant
single and binary star encounters, may also lead to dynamical
instabilities in closely packed multiple planetary systems, resulting
in dissolution of a planetary system or in the merger of some planets
with their host star. The destabilization of these multiple-planet
systems generally occurs through secular and nonlinear interactions
over many orbital periods (Zhou {\it et al.} 2007).

Taking these physical processes into account, there is a need to
investigate the effect of stellar encounters on planetary systems with
more general initial condition than previously considered.
Preliminary studies (see e.g. the different results obtained from
direct $N$-body simulations of Hurley \& Shara 2002) indicate that the
planetary retention efficiency may be strongly modified if the planets
attain a modest eccentricity shortly after their birth. Ideally, the
dynamical interaction between multiple planets and cluster stars needs
to be investigated.  In reality, the vast range of possible orbital
configuration and mass distribution make any attempts to represent
potential planetary systems futile. The orbital periods of compact
planetary systems are much shorter than the crossing time of the
cluster. On the time scale of a large number of orbital periods,
multiple planets interact with each other through cumulative
low-amplitude secular perturbations. The accuracy requirement for
calculating the long-term behaviour of multiple-planet systems is much
higher than that needed for the reliable simulation of star clusters.
For long-term solar system dynamics, symplectic methods, using a
generalized leap-frog, like the widely used Wisdom-Holman symplectic
mapping method (Wisdom \& Holman 1991, see also review by Duncan \&
Quinn 1993), are the best suited integration methods. They do not show
secular errors in energy and angular momentum. In principle, the
symplectic mapping methods can be used to treat the planets, while the
dynamics and perturbation induced by their host stars is computed with
an $N$-body scheme.  However, in their standard implementation, the
symplectic mapping methods require a constant time step, which is not
compatible with the central motivation of the Ahmad-Cohen neighbour
scheme. Another more practical approach to strongly reducing the
planets' secular errors in a $N$-body scheme is to enforce a
time-symmetric scheme by making the time steps reversible through an
iteration (Hut, Makino, \& McMillan 1995, Funato et al. 1996, Kokubo,
Yoshinaga \& Makino 1998). Such schemes have not yet been used for
long-term secular evolution of planetary systems.

Because of these complications, it would be immensely challenging to
consider the effects of secular interactions between closely-spaced
multiple planets concurrently with the dynamical evolution of the
cluster. As an alternative, we may aim to disentangle the different
physical effects, and we begin here by focusing solely on the
influence of gravitational encounters on an otherwise unperturbed
planetary orbit. Having derived cross sections and time scales for
these processes as functions of relevant parameters of the planetary
system and its environment, we will be in a much better position to
assess the conditions under which internal and external perturbations
of planetary systems will couple, and those in which they will act on
very different time scales.

This analysis can also be applied to address the issue of a lack of
planets in globular clusters.  In a recent search for transit events
among stars in the core region of an old, metal-poor globular cluster,
47 Tuc, no short-period planets have been found (Weldrake 2007 and
references therein), even though 17 such objects were expected to be
discovered (Gilliland et al. 2000).  One possible cause for the
absence of planets in globular clusters such as 47 Tuc may be the
suppression of their formation due to lack of heavy elements in their
progenitor clouds (Ida \& Lin 2005a).  However, we note that the mean
metallicity deficiency (compared with the solar value) of 47 Tuc is
smaller than the spread in the dust mass among disks around T Tauri
stars (Beckwith 1998).  The detection of another planet around a
pulsar in globular cluster M4 confirms that although their formation
may be suppressed it is not prohibited
around metal-deficient stars.  However, its stellar
allegiance may have been altered through an exchange/ capture event
during a close encounter between its original parent and its present
host star (Sigurdsson et al. 2003).

Short-period (a few days) planets reside deeply in the potential of
their host stars and can only be perturbed by a very close encounter
between stars. Using an order of magnitude estimate 
for the encounter time scale between a single star and a
planetary system, Bonnell et al. (2001)
 argued that, in globular clusters, planets with
semi major axis greater than 0.3 AU may be detached from their host
stars.  This estimate is based on an extrapolation from the outcomes
of encounters between systems of three bodies of comparable mass,
which may not be appropriate for the limiting case in which one of
these bodies (the planet) is much less massive than the others
\citep{Fr06}.

In dense globular clusters, occasional close stellar encounters can
induce eccentricity excitation even for planets with relatively short
periods (Davies \& Sigurdsson 2001). In multiple-planet systems,
eccentricity excitation of long-period planets can also affect
close-in planets through the secular interaction between them (cf
Mardling \& Lin 2002, Nagasawa \& Lin 2005). In addition, tidal
dissipation inside both short-period planets and their host stars can
damp their eccentricity as indicated by the negligible eccentricity
observed among all planets with period $< 6-7$ days (cf Dobbs-Dixon
{\it et al.} 2004).  The combined influence of these effects heats the
planetary interior and may cause the planet to inflate (Bodenheimer {\it et
al.} 2001).  Very close to their host stars, tidal inflation may be
sufficiently strong to disrupt planets (Gu {\it et al.}  2003). As
the stellar relaxation process leads to an increase in the density of
background stars, their perturbation on their planetary companions
intensifies. This process may be particularly effective in eliminating
planetary companions of relatively low-mass stars in a globular
cluster environment such as 47 Tuc.

In order to 1) verify this induced disruption scenario for the
differences between planetary systems around stars inside and outside stellar
clusters, 2) quantitatively determine the critical condition for
planetary retention, and 3) assess the impact of stellar perturbations
on close-in planets, we consider, in this paper, the dynamical
evolution of planet-bearing stars in a range of cluster
environments. 
In contrast to previous
investigations, we focus on the diverse dynamical structure of
planetary systems rather than the distribution of stellar kinematics,
mass, and multiplicity. For simplicity of analysis, we only consider
interactions between single stars and planetary systems.  In realistic
clusters where their fraction is large, binary stars may indeed
provide more effective perturbers than single stars for the break-up of
planetary systems, especially during the early epoch of cluster
evolution when the stellar density is relatively high. Nevertheless,
interactions between single stars are also important for the long-term
survival of planetary systems in star clusters.

In order to resolve some of these outstanding issues in both open and
globular cluster environments, we carry out a series of numerical
simulations which are designed to investigate the influence of stellar
encounters on the dynamics of extrasolar planets with a wide variety
of orbital properties.  In contrast to previous studies we use a
special $N$-body code (NBODY6++: Spurzem 1999), which is based on the
NBODY6 code by Aarseth (1999a,b, 2003) and can be used on
massively parallel computers.  In addition we use the hybrid Monte
Carlo model (HMC) of Giersz \& Spurzem (2003) as an approximate model
of star clusters with large particle numbers and many planetary
systems.

These simulations, though more time consuming than plain statistical
approaches in which only encounters are modeled, have the advantage
that they can take into account both the spatial and time variation of
the stellar background in young, open, and globular clusters.  In a
previous series of experiments, Davies \& Sigurdsson (2001) have
studied a Monte Carlo model of planetary systems interacting with
stars and binaries in a dense star cluster such as 47 Tuc. However
they did not include the dynamical evolution of the cluster and just
studied isolated three-- and four--body encounters. (Also their
coverage of parameters was smaller than ours.) Both the NBODY6++ and
HMC code are well suited to follow a wide range of encounters and to
overcome some of these shortcomings.

The NBODY6++ scheme is also ideally suited to the inclusion of the
dominant dynamical influence of binary stars and occasional
hierarchical triple systems, but in this paper we only focus on
interactions of single stars with planetary systems.  Another
advantage is the possibility of simulating the consequence of several
successive stellar encounters. After the planetary eccentricities are
slightly excited by the first encounters, their rate of increase may
be accelerated during the subsequent encounters.

In this paper, we first focus on a comparison between empirical cross
sections for orbital changes and analytical estimates (using several
approximations).  This study is intended to clarify the extent to
which the existing analytical cross sections can be used, and we also
are interested to see from the numerical models where they cannot be
used any more, e.g. for the case of liberation of planets, either by
successions of weak encounters or single strong ones. Our main
interest is to understand the physical mechanisms, and an improvement
for more realistic environments (binaries, multi-planetary systems and
their internal interactions) is the subject of future work.  Section~2
summarizes some of the existing knowledge on analytical cross sections
for changes of eccentricity and semi-major axis of a planetary system
due to encounters. In Section~3 we describe the setup of the numerical
simulations and Section~4 contains a description and analysis of the
results.  We pay special attention to the changes of planetary orbits
in a statistical way by searching for encounter events which affect
their eccentricities and semi-major axes (see for comparison also
Theuns 1996).  Once identified, data on these events are accumulated,
and the results are then binned and presented as empirical
differential cross sections for the outcomes of encounters between
single stars and planetary systems; here we use methods developed for
the Monte-Carlo models and those of Giersz \& Spurzem (2003).
Finally, we summarize the findings of these numerical experiments and
discuss their astrophysical implications in Section~5.

\section{Analytical Results on Encounters with Planetary
Systems}\label{sec:analytic_results}

\subsection{General Analytical Approach }

We aim to study the statistics of encounters between a passing star
(the third body) and a simple planetary system consisting of a single
planet and its star.  We focus on the change in eccentricity $\delta
e$ and the relative change in binding energy $\Delta =
\delta\varepsilon/\varepsilon$.  These depend on all the initial
parameters of the encounter, but the essential parameters are the
distance of closest approach, $r_p$, and the speed of the third body
at infinity, $V$ (see Figs.~\ref{fig:scatter-hard} and
\ref{fig:scatter-soft}).  If $r_p\gg a$, where $a$ is the semi-major
axis of the planet, then the encounter is {\sl tidal}.  If $V$ is much
larger than the orbital speed of the planet then the encounter is {\sl
impulsive}, unless $r_p$ is so large that the time scale of the
encounter becomes comparable with the period of the planet.  At still
larger $r_p$ the encounter is {\sl adiabatic}, which means here that
the angular speed of the planet is much larger than that of the
passing star.  If $V$ is small compared to the orbital speed of the
planet, then a tidal encounter is always adiabatic.  When $r_p$ is
very large the path of the passing star is {\sl hyperbolic} with high
eccentricity, but at smaller values (though still in the tidal regime
$r_p\gg a$) it is nearly {\sl parabolic}.  These distinctions are
essential in the analytical interpretation of numerical data.

Appendix A collects a number of formulae for $\Delta$ and $\delta e$
which are approximately valid deep within some of these regimes.  In
adiabatic regimes an important role is played by parameters $e'$ and
$K$, which serve to quantify these ideas.  The first parameter is
just the eccentricity of the relative orbit of the passing star. The
parameter $K$ is defined by 
\begin{equation}
K = \sqrt{{2M_{12}\over M_{123}}\Bigl({r_p \over a }\Bigr)^3}
\end{equation}
where $r_p$ denotes the distance of pericentre for the passing star,
the mass of the binary is $M_{12}=m_1 + m_2$, the mass of the third star
is $m_3$, and $M_{123} = M_{12} + m_3$. In our simulations, $m_1 = m_3$
and $m_2 \ll m_1$, and so we have $K = (r_p / a)^{3/2}$. $K$ measures the
ratio of the time scales involved, and a necessary condition for an
adiabatic encounter in the sense explained above is $K\gg 1$. However,
this criterion by itself would be sufficient only if the encounter
with the third star were parabolic. If the encounter is hyperbolic the
condition for an adiabatic encounter is given by $K/\sqrt{e^\prime + 1}
\gg 1$ (Heggie \& Rasio 1996), in which  the hyperbolic eccentricity
\begin{equation}\label{eq:eprime}
e^\prime = \sqrt{1 + \Bigl({pV^2\over GM_{123}}\Bigr)^2},
\end{equation}
where $p$, $V$ denote the impact parameter and relative velocity at
infinity. With these parameters we get for the minimum distance
$r_p$ of the third star relative to the centre of mass of the
planetary system
\begin{equation}
r_p= {GM_{123}\over V^2} (e' - 1).
\end{equation}
The impact parameter $p$ and distance of closest approach $r_p$ are related by
\begin{equation}
p = r_p \sqrt{1 + {2GM_{123}\over r_p V^2}}.
\label{eq:prp}
\end{equation}

Now we consider the orbital response of a planet as a consequence of
stellar encounters.  Due to perturbations by passing stars, the
semi-major axis and eccentricity of a planet experiences successive
changes $\delta a$ and $\delta e$ which correspond to
changes in its binding energy and angular momentum per unit 
mass, according to the equations
\begin{equation}
\Delta = {\delta E \over E} = - {\delta a \over a}
\label{eq:dE}
\end{equation}
\begin{equation}
{\delta J \over J} = {1 \over 2 } {\delta a \over a} - {e\, \delta e \over 1
- e^2}.
\label{eq:dJ}
\end{equation}
The orbital energy per unit mass $E$ is inversely proportional to
the semi-major axis, and for a circular orbit it is just 
minus half the squared velocity.  Accompanying  the changes in $e$ and $a$ are
changes in $i$, $\omega$, and $\Omega$, but we pay no attention to
these in this paper.

The formulae in Appendix A give approximate expressions for $\Delta$
and $\delta e$ for a single encounter, and depend on all the initial
conditions.  For statistical purposes it is more useful to consider
instead {\sl cross sections}, $\sigma$.  For example $\sigma(\Delta)$,
for $\Delta>0$, would be defined as the cross section for encounters
in which the relative change of binding energy exceeds $\Delta$.  This
would be a function of $V,a$, the masses and (possibly) $e$, though
usually we will consider cross sections averaged over the distribution
of $e$.  Often also we consider {\sl differential cross sections} such
as $d\sigma/d\Delta$.

Cross sections essentially involve the impact parameter $p$, while the
size of the changes $\Delta$ and $\delta e$ is more directly related
to the distance of closest approach, $r_p$.  There are two limiting
cases: (i) strongly hyperbolic encounters without significant
gravitational focusing, where $V^2 \gg GM_{123}/p $, so that $e'\gg
1$, $r_p \gg GM_{123}/V^2$, and hence $p \propto r_p$; and (ii) the
case in which $V^2 \ll GM_{123}/p$, $e' \approx 1$, and so $r_p \ll
GM_{123}/V^2$. In this second case we have strong gravitational
focusing and the second term in equation~(\ref{eq:prp}) above
dominates, so that $p \propto \sqrt{r_p}$.  When $V=0$ the encounter
is parabolic and $e'=1$.

A given planetary system may experience encounters of either type, but which
type dominates depends essentially on $V$ and $a$. We have performed
simulations for two ranges of semi-major axes of planetary orbits, which
we denote for reference as ``hard'' (0.03 to 5 AU) and ``soft'' (3-50
AU), cf. Table 1. As is shown in Figs. 1 and 2 our ``soft'' planets
nearly exclusively experience  encounters where the velocity $V$ of the
encountering star at infinity is larger than the orbital velocity
of the planet $v_{\rm orb}$.
In the case of
 ``hard'' planets (in our sense) we find a large amount of
 encounters (but not all) with the opposite case $ V < v_{\rm orb} $.
 Note that our use of the term ``hard'' and ``soft''              
for planetary systems differs from the definition of \citet{Fr06}. 
 They use the velocity $v_c$ as a threshold, such that
 for $V \approx v_c$ the binding energy of the planetary system is comparable to the
 kinetic energy of the incoming star. Such definition is equivalent
  to what is used for  binaries in star clusters (Heggie 1975), where
 hard binaries have a clear tendency to become harder in close encounters.
 However, in our models, the planetary mass is very small compared
 to the stellar mass ($m_2 \ll m_1$), so we have
 $v_c \approx v_{\rm orb} m_2 / m_1 \ll v_{\rm orb} $.
 An inspection of our Figs. 1 and 2 shows that there are practically no
 encounters with $V < v_c$. Therefore we choose our definition
 of ``hard'' and ``soft'' planets, which we will
 use henceforth without quotation marks, just motivated to   
 distinguish two different initial sets of semi-major axes of
 planetary systems. Indeed, in our models there are effectively
only two regimes.  For our {\sl soft} planets, $V$ typically
exceeds the orbital speed of the planet $ V>  v_{\rm orb} $. 
These encounters are
impulsive if $r_p$ is not too large. They become adiabatic and
hyperbolic if $r_p$ is large enough. On the other hand,
many (but not all) encounters of our {\sl hard} planets fall into
the category $V < v_{\rm orb} $. Here tidal encounters at small
enough $r_p$ are parabolic, but hyperbolic encounters occur at large
enough $r_p$.

\subsection{Cross Sections for the Change of Eccentricity}\label{sec:sigma-D}

\subsubsection{Impulsive encounters}\label{sec:imp}

In the case of soft planets, the most important tidal encounters may
be treated as impulsive.  First we provide an estimate for the form of
the cross section, and then state an accurate result.

Let $\tau$ be the time scale for the encounter, i.e. $\tau = r_p/V_p$,
where $V_p$ is the velocity of the passing star at pericentre. Our
assumption here is that $\tau \siml t_{\rm orb}$, where $ t_{\rm orb}$
is the orbital time of the planetary system, i.e.  we are far from the
fully adiabatic limit. Then we approximate the change of velocity of
the orbiting planet due to the perturbation by the passing star as
\begin{equation}
\delta v \approx \tau \, {\delta\!a }  = {4 Gm_3 a \over V_p
r_p^2}
\end{equation}
where $\delta a$ is the difference between the acceleration of the
planet (by the perturber) and that of the parent star (by the
perturber), which we estimate by a tidal approximation.

Now the {\sl relative} changes in { the binding energy},
$\varepsilon$, and angular momentum, $J$, of the planetary 
system may be estimated by $\delta v/v$, and so by
equation~(\ref{eq:dJ}) this may also be used to estimate $D$, which we
define as the change in the {\sl square} of the eccentricity, $D =
\delta(e^2)$.  (This is more convenient than $\delta e$ itself in some
later analysis.)  So we have
\begin{equation}\label{D-estimate}
D \sim {\delta v \over v} \approx \sqrt{Gm_3\over aV_p^2}
\cdot \left({a\over r_p}\right)^2.
\end{equation}
For the last step above we have used $m_3 \approx m_1$. We also use
$V_p^2 = V^2 + 2 G (m_1+m_3)/ r_p $.  Thus
\begin{equation}
D \approx \sqrt{Gm_3\over aV_p^2}  \cdot
\left({a\over r_p}\right)^2 \propto r_p^{-2} .
\end{equation}
The total cross section is $\sigma = \pi p^2 \simeq \pi r_p^2 \propto
D^{-1}$, and
so we get for the differential cross
\begin{equation}
{d\sigma \over dD}\Bigr\vert_{}  \propto 
D^{-2} 
\label{eq:impulsive}
\end{equation}

For a proper estimate we may adopt the 
approach which was
taken by Heggie (1975) for the computation of the cross section for
the relative energy change.  As shown  in Appendix~\ref{sec:imp_tid},
the result is that 
\begin{equation}\label{eq:impulsive-D}
	\frac{d\sigma}{dD} = {2C_1}{}\frac{(Gma^3)^{1/2}}{VD^2},
\end{equation}
where we have specialized to the case in which the two stars (the
passing star and the sun in the planetary system) have the same mass
$m$, and $C_1$ is a constant which is defined in terms of a certain
average, and evaluates numerically to $C_1 = 0.883$ approximately.

\subsubsection{Adiabatic encounters}\label{sec:D_adiabatic}

Now we consider adiabatic tidal encounters.  Using a first-order
expansion, Heggie (1975) and Heggie \& Rasio (1996) obtained the net
secular (i.e.  long-term relative to the binary's orbital
period) changes of eccentricity $\delta e$ for parabolic and
hyperbolic encounters between a single field star and a close binary
star.  We adapt their formula to our case, in which the masses $m_1$
of the planet's host star and $m_3$ of the approaching star are large
compared to the planetary mass $m_2$; so we have $M_{12}\equiv m_1+m_2
\approx m_1$ and $M_{123} \approx m_1 + m_3$.  equation (7) of Heggie
\& Rasio (1996) yields for this case
\begin{equation}
\delta e = - {15 \over 4} \left(m_3^2 \over m_1 M_{123} \right)^{1/2}
\left( {a \over r_p} \right)^{3/2}
{e \sqrt{ 1 - e^2} \over (1+e')^{3/2} } \quad g (e^\prime, \Omega, \omega, i),
\label{eq:de0}
\end{equation}
where $g$ is the function in curly brackets in
eq.(\ref{eq:de-noncirc}) in the present paper.
It reduces to $g=\pi \sin(2\Omega) \sin^2 i$ for parabolic encounters, 
so to order of magnitude we use $g\approx 1$ in this case. For highly 
hyperbolic encounters we have 
similarly $g \approx e'$, $f_3(e') = e'/(1 + e')^{3/2} \approx e'^{-0.5} $, 
and we use $e^\prime\propto p \approx r_p \propto K^{2/3}$. It follows
that
\begin{eqnarray}
\delta e_{\rm par} &\propto & K^{-1} \propto r_p^{-3/2} \cr
\delta e_{\rm hyp} &\propto & K^{-4/3} \propto r_p^{-2}
\label{eq:de}
\end{eqnarray}
From equations~(\ref{eq:de}) we can deduce approximate total cross
sections for both limits, using $\sigma\propto p^2 \propto r_p^2$ for
the hyperbolic case, and $\sigma\propto p^2 \propto r_p $ for the
parabolic case. It follows that
\begin{eqnarray}
\sigma_{\rm par} &\propto & (\delta e)^{-2/3} \cr
\sigma_{\rm hyp} &\propto & (\delta e)^{-1}
\end{eqnarray}
and so we recover in the parabolic case the form of the result in Heggie
\& Rasio (1996) (see below). 
From this, differential cross sections are computed by
\begin{eqnarray}
{d\sigma\over d (\delta e)}\Bigr\vert_{\rm par} &\propto & (\delta e)^{-5/3}\cr
{d\sigma\over d (\delta e)}\Bigr\vert_{\rm hyp} &\propto & (\delta e)^{-2}
\label{eq:cross-de}
\end{eqnarray}

The details are given in Appendix B, where the following accurate
results are obtained.  (Note, however, that these are differential
cross sections for the change in $e^2$, i.e. $D = \delta(e^2)$.)  In
the case of extremely hyperbolic encounters with hard planets, the
calculation is carried out in Appendix~\ref{sec:D_hyperbolic}, and
yields a formula of the same form as for impulsive encounters:
\begin{equation}
	\frac{d\sigma}{dD} = \frac{16C_2}{3}\frac{\sqrt{Gma^3}}{VD^2},
\end{equation}
where $C_2 = 0.5932$ approximately.  For parabolic encounters,
however, the corresponding result was essentially given by equation
(19) of \citet{HR96}.  As shown in Appendix~\ref{sec:D_parabolic} this
leads to the differential cross section
\begin{equation}
	\frac{d\sigma}{dD} = \frac{2}{21}(15\pi)^{2/3} 
\left[\Gamma\left(\frac{2}{3}\right)\Gamma\left(
\frac{5}{6}\right)\right]^2\frac{Gma}{V^2}(D)^{-5/3}.
\end{equation}
The form of this cross section is different, because of gravitational
focusing.

\subsection{Cross sections for the change in energy}

\subsubsection{Impulsive encounters}\label{sec:impulse}

We consider now the relative change in energy, $\Delta$, and begin
with impulsive tidal encounters.  
As already mentioned, formulae were given by Section~4.2 of \citet{H75} for the
change in energy in an impulsive tidal encounter, and all that is
needed is to specialize to the case $m_1 = m_3 = m$ and $m_2\to0$, and
to harmonize the notation.  He gives (his equation~(4.24)) a cross section
for the change $y$ in binding energy $x$.  In the language of the
present paper his cross section is our differential cross section
$d\sigma/d\Delta $.  With $\Delta = \delta\varepsilon / \varepsilon$ we readily find
that 
\begin{equation}\label{eq:Delta_impulsive}
	\frac{d\sigma}{d\Delta} = \frac{2\pi}{3}\frac{\sqrt{Gma^3}}{V\Delta^2},
\end{equation}
though Heggie used units in which $G=1$ and so we have restored it in
its proper place here.

This formula makes no distinction of the sign of $\Delta$: it appears
that hardening and softening encounters are equally likely.  But
Section~4.3 of \citet{H75} also showed that an application of the principle
of detailed balance could be used to estimate the difference in the
{\sl rate} of hardening and softening encounters.  A similar
principle, however, applies to differential cross sections, and we
employ it here.  From equation~(A3) in
\citet{HH93}, which is expressed in slightly different
notation, it is easy to see that
\begin{equation}
	\frac{d\sigma}{d\Delta}(\Delta,a,V)a^{7/2}V^2 = 
	\frac{d\sigma}{d\Delta}(\Delta',a',V')a'^{7/2}V'^2,
\end{equation}
where we have made explicit the dependence of the cross section on the
initial semi-major axis of the planet and the initial speed of the
passing star.  On the left side we are here considering encounters
which begin with values $a,V$ and end with values $a' = a/(1+\Delta)$
and $V'$.  The right side considers information on the time-reversal
of such encounters, and so $(1+\Delta)(1+\Delta') = 1$.  
In the present case, with $m_2 = 0$, we have $V =
V'$.  Substituting for $a'$ and $V'$ we deduce that
\begin{equation}
	\frac{d\sigma}{d\Delta}(\Delta,a,V) = 
(1+\Delta)^{-7/2}  \frac{d\sigma}{d\Delta}(-
\frac{\Delta}{1+\Delta},a/(1+\Delta),V).
\end{equation}
Substituting from equation~(\ref{eq:Delta_impulsive}) we deduce that
\begin{equation}
\frac{d\sigma}{d\Delta}(\Delta,a,V) = 
(1+\Delta)^{-3}  \frac{d\sigma}{d\Delta}(-{\Delta},a,V).  
\end{equation}
Finally, expanding for
small $\vert\Delta\vert$, we find that
\begin{equation}
\frac{d\sigma}{d\Delta}(\Delta,a,V) - \frac{d\sigma}{d\Delta}(-\Delta,a,V) \simeq -
3\Delta  \frac{d\sigma}{d\Delta}(\Delta,a,V).\label{eq:difference}
\end{equation}

This interesting result shows that hardening encounters 
are rarer than softening ones.  The same qualitative conclusion results from
considering non-tidal impulsive encounters, i.e. impulsive encounters
in which the passing star approaches the planet or its sun to a
distance less than $a$.  For completeness we present here two results
for this case drawn from equation (4.12) in \citet{H75}, but specialized to the present
masses and translated into the notation of the present paper.  They are
\begin{equation}\label{eq:Delta_non-tidal}
	\frac{d\sigma}{d\Delta} = \frac{8\pi Gma}{V^2\Delta^2}\left\{
	\begin{array}{cc}
\displaystyle{ \left(1-\frac{4}{3\Delta}\right) }\ \ \ \ \ ,&{\Delta<0}\\
\displaystyle{\left(\frac{7}{3} + \frac{4}{3\Delta}\right)
\left( 1+\Delta \right)^{-5/2}},&{\Delta>0}.    
	\end{array}\right.
\end{equation}
Again we find that softening encounters have a higher cross section
than hardening encounters.

Though these cross sections apply to impulsive encounters, which is a
regime of 
relevance for soft planets, it was pointed out by Section~5.1 of \citet{H75}
that the results should be roughly applicable also for non-tidal
encounters with hard systems, except for a correction due to
gravitational focusing.  The passing star is accelerated by the
star of the planetary system until its relative velocity becomes
comparable with that of the planet.  In the case of stellar-mass
binaries, which was the application in \citet{H75}, the outcome for
hardening encounters ($\Delta>0$) is complicated by the possible capture of the
passing star.  This does not occur in the case of a planetary system
in the limit $m_2\to0$, and so equation~(\ref{eq:Delta_non-tidal}) does
apply roughly (with the aforementioned correction) to planetary
systems.  It also helps to explain the discovery by \citet{Fr06} that
a planetary system of intermediate binding energy softens on average (provided that it is not so
hard that capture of the passing star is energetically possible). Intermediate
refers here to a planetary system in the regime $v_c < V < v_{\rm orb}$, where
$v_c$ is the critical velocity according to \citet{Fr06} and $v_{\rm orb}$ is the
orbital velocity of the planet. 

\subsubsection{Adiabatic encounters}

We can discuss changes of the binding energy 
in much the same way that we discussed changes in eccentricity in
Section~\ref{sec:D_adiabatic}.  For the parabolic case, expressions for
$\Delta$ are given by \citet{H75} and, more explicitly, by 
Roy \&
Haddow (2003), while  Heggie (2005) extends these to the
hyperbolic case.  Again we here give only enough  to explain the form of
the cross sections, and defer further detail to the Appendix.

Let $\varepsilon$ denote the binding energy of a
planetary system, with semi-major axis $a =
Gm_1m_2/2\vert\varepsilon\vert$, and $m_2 \ll m_1$.  Then the
relative binding energy change $\Delta = \delta\varepsilon /
\varepsilon$ for an encounter becomes
\begin{equation}
	\Delta = - \sqrt{\pi} f_1(e') K^{1/2} \exp\left( -{2\over 3}K f_2(e')\right)
	F(e,\omega,\Omega, nt_0)
\label{eq:Delta}
\end{equation}
The definition of $F(e,\omega,\Omega, nt_0)$ can be obtained by comparison with
equation~(\ref{eq:deps}). This factor is obtained from the one given by Roy \& Haddow
(2003), equation~(19) and
Heggie (2005), equation~(11) by taking out the factor $a^2$.
Thus our function $F$ is of order unity, and depends only on the
eccentricity,  orientation and
phase of the planetary orbit. For the purpose of
our paper only
an average over all possible values is of interest, and to order of magnitude we use here
$F\approx 1$.
Now we quote the two functions
\begin{eqnarray}
	f_1(e') &=& \left({ e'+1 \over 2} \right)^{3/4} e'^{-2} \cr
	f_2(e') &=& {3\over 2\sqrt{2}} \, { \sqrt{e'^2 - 1} - \arccos(1/e') \over
	(e'-1)^{3/2} }.
\label{eq:f1f2}
\end{eqnarray}
We discuss the functions for the parabolic case $e'=1$ and for the extremely
hyperbolic case $e'\gg 1$; for the first one we have
$f_1(e')=f_2(e')=1$, and in this case we reproduce the result of Roy
\& Haddow (2003). In the hyperbolic limit we have asymptotically
$f_1(e') \approx e'^{-5/4}$ and $f_2(e') \approx e'^{-1/2}$, and  so we
get
\begin{eqnarray}
\Delta_{\rm par} &\approx& - \sqrt{\pi}  K^{1/2} \exp\left( -{2\over
3}K \right) \cr
\Delta_{\rm hyp} &\approx& - \sqrt{\pi} e'^{-5/4}
K^{1/2} \exp\left( -{2\over 3}K e'^{-1/2}\right).\label{eq:Delta_hyerbolic}
\end{eqnarray}
As before we have  $e' \propto p \approx r_p$ for the hyperbolic case,
and in both cases $K\propto r_p^{3/2}$ by definition, whence it
follows that
\begin{eqnarray}
\Delta_{\rm par} &\propto& - \sqrt{\pi}  r_p^{3/4} \exp\left(
-{2\over 3} r_p^{3/2}\right) \cr
\Delta_{\rm hyp} &\propto& -
\sqrt{\pi} r_p^{-1/2} \exp\left( -{2\over 3} r_p \right).\label{eq:Delta_hyp}
\end{eqnarray}
To compute differential cross sections we need to invert the
function $\Delta(r_p)$, which can be done easily only if the
exponential function dominates. In that case we have
\begin{eqnarray}
\sigma_{\rm par} &\propto & p^2 \propto r_p \propto (\ln\Delta)^{2/3}  \cr
\sigma_{\rm hyp} &\propto & p^2 \propto r_p^2 \propto (\ln\Delta)^2,
\end{eqnarray}
where we think of $\Delta$ as positive.
Thus differential cross sections are obtained in the forms
\begin{eqnarray}
{d\sigma \over d\Delta}\Bigr\vert_{\rm par}  &\propto & {1\over \Delta }
(\ln\Delta)^{-1/3} \cr
{d\sigma \over d\Delta}\Bigr\vert_{\rm hyp}  &\propto &  {1\over \Delta } \ln\Delta.
\label{eq:crossdelta}
\end{eqnarray}

It is shown in Appendix~\ref{sec:Delta_cross_sections} that, to
leading order, the full results are
\begin{eqnarray}
{d\sigma \over d\Delta}\Bigr\vert_{\rm par}  &= & \frac{2\pi Gma}{V^2}{1\over
\Delta } (\ln\Delta)^{-1/3} \cr
{d\sigma \over d\Delta}\Bigr\vert_{\rm hyp}  &= & \frac{\pi a^3V^2}{Gm} {1\over
\Delta } \ln\Delta.
\label{eq:crossdelta_full}
\end{eqnarray}

\section{Computational Methods and Initial Model Parameters}

We use our hybrid Monte Carlo (HMC) scheme as well as direct $N$-body
simulations to study the changes in the orbital elements of planetary
systems induced by encounters in stellar clusters.  The main
limitations of the models are that all stars are single (except for
the planetary companions), all have equal mass, and there is no
stellar evolution.  The main loss of generality is the choice of a
specific range of semi-major axis for the planetary orbits.  Though we
argue below that the range adopted is astrophysically justified, it is
one aim of the analytical work to show how the results might be
generalized to values outside this range.  We collect data for
encounters between stars and planetary systems in different ways in
the two models, as described in the following sections.  The initial
parameters of $N$-body and HMC models are summarized in
Table~\ref{tab:constants}.

\subsection{Direct $N$-Body models}\label{sec:nb}

The N-body code we use is the publicly available NBODY6++ version.
The equations of motion of each planetary system can be treated by
regularization.  Encounters and perturbations by flybys affect the
orbital elements of these regularized pairs, and a sufficiently strong
interaction with a passing star can dynamically dissolve the planetary
systems.  The only special refinement to mention is that we utilize
the newly applied classical method with which secular errors in the
integration of close binaries in stellar systems can be strongly
reduced (Mikkola \& Aarseth 1998).

For all models, we adopt an isotropic Plummer model for the stars'
initial phase space distribution function.  This model provides a reasonable
approximation for open cluster potentials.
All models are in dynamical
equilibrium initially.
Our model units are such that $G=1$, $M=1$, $E=-0.25$, for the
gravitational constant, initial total mass and energy, respectively
(standard $N$-body units, \citet{HM1986}). For all models, the individual mass of
stars
thus scales with $1/N$. Planetary masses are set to the constant value of 
$10^{-10}$ in $N$-body units for both the $N$-body and HMC models. This choice
ensures that all our planets are practically massless compared to the stellar
masses.
Physical units are obtained by (a) assigning an
individual
stellar mass ($m_\ast$) in units of solar masses (see Table~\ref{tab:constants}) to
the stars
and (b) defining 1 pc as one $N$-body unit (this is the virial radius $R=GM/4\vert E\vert $).
The length scaling law also fixes what is one $N$-body unit in
AU ($2.27 \cdot 10^5 $ for a system with solar-mass stars and 1 pc
identified with one $N$-body unit).

For all runs done here all stars have equal mass, and there are no stellar
binaries initially.  We place one and only one planet around $N_p$ of
the stars. For the direct $N$-body models we use 20,000 objects, $N_p=1000$
of which are planets, which are randomly attached to 19,000 equal mass
single stars, to form a planetary system. This is a smaller particle
number than in the HMC runs (see below).  Other properties,
such as the initial Plummer model, and the absence of tidal fields, are
exactly as in the case of Monte Carlo models.

In all of our $N$-body runs except two (runs E, EH) all planets have
the same initial eccentricity $e_0$, and we do different runs with
differing values of $e_0$ (Models 1-6: see Table~\ref{tab:constants}).
The eccentricities of our models range from 0.01 (in model 1) to 0.99
(in model 6).  The use of several different models with different
eccentricities has two effects: (i) it improves the statistical data
for the relatively small particle number (and relatively small number
of scattering events) in the direct $N$-body simulations; (ii) it
allows us to consider the dependence of scattering statistics on the
planet's orbital eccentricity, and to facilitate a more
straightforward comparison with analytic models (Heggie \& Rasio 1996,
Roy \& Haddow 2003, Heggie 2005). Runs E and EH, however, cover all
initial eccentricities using a thermal eccentricity distribution $f(e)
=2e$, which is also adopted in the HMC models. For small $e$'s, this
distribution approximates the Rayleigh distribution which is expected
as a consequence of dynamical instability and relaxation (Zhou {\it et
al.} 2007).  In all calculations, the initial phase and orientation of
the planets' orbits, including the direction of their angular momentum
vector and their periastron, are randomized in the obvious senses.

The semi major axes of the planets are chosen with a logarithmic
distribution, i.e. a constant $d N_p / d {\rm log} a$, between 3
-- 50 AU for run E and between 0.03 -- 5 AU for run EH.  The initial
distribution of semi-major axes in the runs E and EH corresponds to
the soft and hard planetary systems in the HMC runs, respectively.
Note that we denote these planetary systems as soft and hard
to distinguish between the two types of runs (with physically wider
and closer planetary systems). 
We would like to remind the reader again here that our definition of hard and soft
differs from the dynamical definition of \citet{Fr06}. According to the dynamical
definition a planetary system is hard if $V < 0.75 v_c$, where $v_c$ is the critical
velocity of the three-body system. We choose the orbital velocity $v_{\rm orb}$ of
the planetary system as parameter to distinguish encounters.
If we use for the average $V^2$ in our models twice the
cluster velocity dispersion, we get in $N$-body units $V^2/v_{\rm orb}^2 = N
(a_{\rm AU}/2.27\cdot 10^5)$, where we have again taken one $N$-body
length unit to be 1pc and $a_{AU}$ is the planet's semi major axes in
astronomical units. In Table~\ref{tab:constants} we give the value of
$V/v_{\rm orb}$ for our initial planetary systems.

In choosing these parameters we have been guided by theoretical and
observational considerations.  The range of 3-50 AU of semi major
axis represents the location where gas giant planets are most likely
to form (Ida \& Lin 2005).  Migration due to protoplanet-disk
interaction may repopulate the regions interior to 3 AU (Lin,
Bodenheimer \& Richardson 1996) and dynamical instabilities could
eject planets beyond 50 AU (Lin \& Ida 1997).  But most of the gas
giant planets may remain near the location of their formation.  While
gas giants are most likely to have formed with nearly circular orbits,
dynamical instabilities could excite their eccentricities to the point
of ejection.

Up to now, one or a few Jupiter-mass planets, with periods less than a
few years, have been found around less than 10\% of the nearby solar-type
stars (Marcy et al. 2005), though planets with longer periods are
expected to be more common (Trilling et al.  1998, Armitage et
al. 2002, Cumming {\it et al.} 2007). The total mass of planets is too
small to significantly perturb the internal dynamics of any stellar
cluster.  Thus, for the simulations to be presented here, we include
$N_p$ planets with very small masses of order $10^{-10}$, such that
the mass ratio of planetary to average stellar mass is $1.9\cdot 10^{-6}$,
comparable to the mass of terrestrial planets.
The total and individual mass of our planets is small enough, that they do
not contribute any significant dynamical feedback to the cluster of
$N_s$ stars.

As for the stellar system itself,  our aim has been to represent a
typical rich young star cluster such as the Orion region. However,
most  stars are formed in binary and multiple systems. In stellar
clusters, the presence of binary stars can significantly speed up the
relaxation process due to their larger mass (Gao et al. 1991). They
also strongly enhance the frequency of close three- and four-body
encounters due to their larger cross section. This will also affect
planetary systems (Laughlin \& Adams 1998), and the influence of
interactions with binary stars on planetary systems will be
investigated in future work.

\begin{table}[t!]
\caption{Initial parameters of  $N$-body and HMC models $^{\rm a}$}
\begin{tabular}{cc|cccccc} \hline\hline
Model & $N_\ast$ & $m_\ast (M_\odot)$ & $N_p$ & $a$(AU) & $e_0$ &
$V/v_{\rm orb}$~$^{\rm b}$ & Type \\ 
\hline 1 & $1.9\cdot 10^4$ & 1 & $10^3$ & 3-50 & 0.01 & 0.49 - 2.05 &
$N$-body \\ 
2 & $1.9\cdot 10^4$ & 1 & $10^3$ & 3-50 & 0.1 & 0.49 - 2.05 & $N$-body
\\ 
3 & $1.9\cdot 10^4$ & 1 & $10^3$ & 3-50 & 0.3 & 0.49 - 2.05 & $N$-body
\\ 
4 & $1.9\cdot 10^4$ & 1 & $10^3$ & 3-50 & 0.6 & 0.49 - 2.05 & $N$-body
\\ 
5 & $1.9\cdot 10^4$ & 1 & $10^3$ & 3-50 & 0.9 & 0.49 - 2.05 & $N$-body
\\ 
6 & $1.9\cdot 10^4$ & 1 & $10^3$ & 3-50 & 0.99 & 0.49 - 2.05 &
$N$-body \\ 
E & $1.9\cdot 10^4$ & 1 & $10^3$ & 3-50 & $f(e)=2e$ & 0.49 - 2.05 &
$N$-body \\ 
EH & $1.9\cdot 10^4$ & 1 & $10^3$ & 0.03-5 & $f(e)=2e$ & 0.05 - 0.65 &
$N$-body \\ 
Soft & $3.0\cdot 10^5$ & 1 & $3.0\cdot 10^4$ & 3-50 & $f(e)=2e$ & 1.98
- 8.13 & HMC \\ 
Hard & $3.0\cdot 10^5$ & 1 & $3.0\cdot 10^4$ & 0.03-5 & $f(e)=2e$ &
0.20 - 2.57 & HMC \\ 
\noalign{
\parbox{0.9\linewidth}{\bigskip $^{\rm a}$ $N_\ast$ is the total
initial number of stars and planetary systems, $m_\ast (M_\odot)$ is
the mass of one star in solar units, $N_p$ is the initial number of
planetary systems, $a$ (AU) is the initial range of the semi-major
axes in AU, and $e_0$ is the initial eccentricity.\par\noindent $^{\rm
b}$ we use $V/v_{\rm orb} = \sqrt{N (a_{\rm AU}/2.27\cdot 10^5)}$ 
to obtain an approximate range of values, cf. Section 3.1.
\bigskip}}
\end{tabular}
\label{tab:constants}
\end{table}
\eject

There is one important technical aspect of these $N$-body runs that
remains to be described.  While direct $N$-body models contain less
intrinsic approximations than other simplified models, such as the HMC
model, for our purposes there is a drawback, because it is very
difficult to identify isolated two-body encounters in an $N$-body
model. In fact it has been discussed, whether a real $N$-body system's
relaxation process can be described by the standard model of
uncorrelated small angle two-body encounters between individual stars
(Theuns 1996). Generally, it is possible to identify an encounter by checking the
minimum distance to the closest neighbour of any given particle, to
get the $r_p$ and the velocity $V_p$ at closest distance.  However, it
is very difficult to determine the proper initial parameters of an
encounter, because the scintillation and fluctuation of the $N$-body
potential perturbs any orbit even at moderate distances. Despite 
all these factors  it is possible operationally to
determine encounter event data, similar to that and
to be compared with the HMC model. In fixed
time intervals of one $N$-body time unit (approximately one half-mass
initial crossing time) we monitor orbital elements of all planetary
systems ($e$, $a$). If any one of them has changed by more than
$5\cdot 10^{-7}$ since the previous time, we assume an encounter took place.
We measure $\delta e$ and $\Delta$, 
and thus have a data bank of encounters for the $N$-body
system similarly to that for the HMC runs. From a theoretical point of view
there may be uncertainties about this procedure, but we don't know any
better alternatives, and judging from the results it seems to be a
reasonable operational procedure. The value of the hyperbolic
eccentricity $e'$ cannot be determined a priori; however from
comparison with analytical models (see figures below) one can see that
the  deduced values of $e'$ lie in a completely reasonable range.

\subsection{Hybrid Monte Carlo method}

We use the hybrid Monte Carlo (HMC) method developed by Spurzem \&
Giersz (1996), Giersz \& Spurzem (2000, 2003) to model the evolution
of a star cluster with a large number of stars and planetary
systems. The latter are considered as if they were a binary; binaries
are treated with the Monte Carlo scheme to follow their relaxation
with each other and with single stars in the cluster, while single
stars are described by the anisotropic gaseous model based on the
Fokker-Planck approximation (Louis \& Spurzem 1991).  Close encounters
between planetary systems and a single star are followed as in the
cited papers using a direct few-body integrator employing
regularization methods (cf. e.g. Mikkola 1997). 
For HMC the planet mass in N-body units
is very small, $6.33E-12$ and the mass ratio of planets to stars
is 1.9E-6 also, as in the
case of $N$-body runs, and consistent with the notion that our planets have
a mass comparable to terrestrial planets. 

We describe results of two sets of models, each with 300,000 single
stars, 30,000 of which initially have a planet (see Table~
\ref{tab:constants}). The semi-major axes of planetary systems
introduce another independent scale into the otherwise scale-free $N$-body 
system. We have chosen initially to keep the planetary scale $a$ (semimajor
axes) constant relative to the scale radius of the stellar system,
independently of $N$. As a consequence the size of planetary systems
measured in $N$-body units is the same for HMC and $N$-body models. Note
that this means the squared orbital velocity of planets scales as $1/N$,
where $N$ is the particle number. In terms of cluster velocity dispersion
our planetary systems become weaker (softer) with larger $N$. We will discuss
the scaling behaviour of encounters between planetary systems and single stars
in more detail in sections 5.3 and 5.4.

All parameter choices for the planetary systems are
analogous to those of the $N$-body simulations, as explained before,
and some more details are given in the table. For each set of
parameters, two models with independently generated initial phase
space distributions are adopted to boost the statistical significance
of our results.  The runs were continued for approximately five
initial half-mass crossing times.
 This is a small interval for relaxation, but already enough to sample interactions between
 planetary systems and single stars appropriately. The time step of the HMC code is small
 enough to resolve these very accurately in time, because it is given by the gaseous model
 code for the single star component. Despite of small time steps it is taken care that every
 binary is treated by the Monte Carlo procedure for relaxation with the correct rate and
 no over-relaxation occurs. A detailed description of this and the initial setup for
 encounters is given in \citet{Giersz03}. 
Here, we will only summarize the
main points of the setup procedure. Whether a planetary system
actually suffers from an encounter with a star is determined randomly
for each time-step and planetary system. First, a maximum impact
parameter is chosen to cover all physically interesting cases, $p_{\rm
max} = 2000 $ AU. From $p_{\rm max}$ we determine an encounter rate
$\dot{N}$ using the simple prescription $\dot{N}=nAv_{\rm rel}$, where
$n$, $A=\pi p_{\rm max}^2$, $v_{\rm rel}$ are the local planetary
number density, the geometrical cross section corresponding to $p_{\rm max}$, and
the actual relative velocity of the star and planetary system chosen
for an encounter. In every time step we determine whether an encounter
takes place by comparing the probability obtained from the encounter
rate with a random number.  After we have identified an encounter we
pick an actual impact parameter $p$, from a random distribution in
$p^2$.  Using this with energy and momentum conservation we transform
the encounter from the cluster frame into the interaction frame and
start the actual direct integration of the encounter using packages
from Aarseth's NBODY6 code. After the termination of the direct
integration of the encounter the new parameters of the planetary
system are computed and the outcome of the encounter is identified:
flyby, dissolution or exchange and transformed from the interaction
frame into the cluster frame.

\section{Results of numerical experiments}

The three parts of this section summarize our numerical results and
their interpretation.  We begin with a discussion of the destruction
of planetary systems in stellar encounters, and the escape of the
resulting population of freely-floating planets.  Then we give a
largely empirical description of the results of the non-destructive
encounters.  This analysis paves the way for the final subsection, in
which we interpret the results as directly as possible in terms of
differential cross sections.

\begin{table}[t!]
\caption{Summary of results of hybrid Monte Carlo (HMC) runs}
\begin{tabular}{|l|rrrr|} \hline\hline
 Model               \qquad         &  HMC soft 1 \qquad & 2   \qquad      &  HMC hard 1 \qquad & 2\qquad \\
\hline
$N_{\rm events}$                    & 1995262     & 2426598  & 3406206     & 3650990     \\
$t_{N-{\rm body}}$                  & 15.28       &   15.68  & 20.67       &  30.31      \\
$\tau_{\rm cr} = t/t_{{\rm cr},0}$  & 5.40        &   5.54   & 7.30        &  10.71      \\
$\tau_{\rm rh} = t/t_{{\rm rh},0}$  & 5.93E-03    &  6.08E-03& 8.02E-03    &  1.18E-02   \\
$N_{\rm pl-diss}$                   & 149         &  162     & 7           &  9          \\
$N_{\rm pl-diss-esc}$               & 10          &  10      & 3           &  3          \\
$N_{\rm pl-ff}$                     & 139         &  152     & 4           &  6          \\
$N_{\rm pl-ff}/\tau_{\rm rh}$     & 23440       & 25000    & 498.1       & 508.5       \\
$N_{\rm pl-ff}/\tau_{\rm cr}$     & 25.74       & 27.44    & 0.548       & 0.560       \\
$x_{\rm pl-ff,rh}$                & 0.7813      & 0.8333   & 0.0166      & 0.0170      \\
$x_{\rm pl-ff,cr}$                & 8.58E-04    & 9.15E-04 & 1.83E-05    & 1.87E-05    \\
Scaled $x_{\rm pl-ff,rh}$         & 0.0673      & 0.0718   & 1.43E-03    & 1.46E-03    \\
\hline
\noalign{ \parbox{0.9\linewidth}{\smallskip\small
$N_{\rm events}$ is the number of interactions with planetary
systems,
$t_{N-{\rm body}}$ is the time of termination of the
simulation in $N$-body units,
$t_{cr,0}$ is the initial crossing time,
$t_{rh,0}$ is the initial half-mass relaxation time,
$N_{\rm pl-diss}$ is the number of dissolved planetary systems,
$N_{\rm pl-diss-esc}$ is the number of planets escaped from
the system after the dissolution of the planetary system,
$N_{\rm pl-ff}$ is the number of ``freely floating" planets.
\newline
The quantities $x_{\rm pl-ff,rh}$ and $x_{\rm pl-ff,cr}$ denote the
probability for one planet to become a free floater, and are just
obtained from the previous two lines by dividing by the total
number of planets. $x_{\rm pl-ff,cr}$ can be directly compared with
$N$-body results below, compare Eq.~\ref{eq:cross}. $x_{\rm pl-ff,rh}$
will scale with the two-body relaxation time and
therefore as $N_2\ln(\gamma N_1) / N_1\ln(\gamma N_2)$, with $N_1 = 300.000$ (HMC
run),
$N_2 = 19.000$ ($N$-body run), $\gamma = 0.11$, cf. \citet{GierszH94}.
}}
\hline
\end{tabular}
\label{tab:HMCdata}
\end{table}

\begin{table}[t!]
\caption{Summary of results of direct $N$-body runs}
\begin{tabular}{|l|rrrrrrrr|} \hline\hline
	Model &1  &  2     &    3     &    4   &  5     &      6      &     
E       &     EH   \\ \hline
$N_{\rm events}$      &   24338& 45071 &  52060      &    70151 & 86951   &  81655
&   62022      &   6263 \\
$t_{N-{\rm body}}$  & 72.0  &  130.  &      153.     &    170. &  181.  &      166.  
&    134.      &   200.  \\
$\tau_{\rm cr} $        & 25.5  &   46.0 &     54.1      &    60.1 &   64.0  &      58.7 
&     47.4     &   70.7 \\
$\tau_{\rm rh} $        & 0.32  &  0.59  &     0.69      &    0.77 &   0.81  &     0.75  
&     0.60     &   0.9   \\
$N_{\rm pl-diss}$   & 33    &  66    &     69        &    89   &   63    &     63    
&     70       &   2     \\
$N_{\rm pl-diss-esc}$ &  3  &   5    &      4        &    5    &    8    &      8    
&      3       &   0   \\
$N_{\rm pl-ff}$       & 30  &  61    &     65        &    84   &   55    &     55    
&     67       &   2   \\
$N_{\rm pl-ff}/\tau_{\rm rh}$ & 93.8&  103.4&  94.2       &    109.1 &  67.9   &    73.3   
&    111.7     &   2.22\\
$N_{\rm pl-ff}/\tau_{\rm cr}$ & 1.18&  1.33 &  1.20       &    1.39  & 0.85    &     0.94  
&    1.41      &   0.028 \\
$x_{\rm pl-ff,cr}$ &  \hspace{-7pt}    1.18E-03 & \hspace{-7pt}1.33E-03 &  \hspace{-7pt}1.20E-03 &  \hspace{-7pt}1.39E-03 &  \hspace{-7pt}8.5E-04 &  \hspace{-7pt}9.4E-04
&   \hspace{-7pt} 1.41E-03  & \hspace{-7pt}  2.80E-05 \\
$x_{\rm pl-ff,rh}$ &           0.0938 & 0.103 & 0.0942   &    0.109 &  0.0679  &  0.0733 
&    0.112     &  \hspace{-7pt} 2.22E-03 \\
$\xi_{\rm pl-ff,cr}$ &        1.35        &  1.48    &   1.35   &  1.57    & 0.98    & 1.05   
& 1.61         &  1.54     \\
$\xi_{\rm pl-ff,rh}$ &        1.33      &  1.50 &  1.35    &  1.57    &  0.96    &  1.06  
&   1.59       &  1.51     \\
\hline
\noalign{ \parbox{1.0\linewidth}{\smallskip\small
All quantities have the same meaning as in
Table~\ref{tab:HMCdata}. The last two lines give $\xi$, defined as the ratio
of $x$ obtained from the $N$-body model divided by the average of the two
corresponding HMC results, i.e. $\xi_{\rm pl-ff,cr} = x_{\rm pl-ff,cr,Nbody}/x_{\rm pl-ff,cr,HMC,av}$
and $\xi_{\rm pl-ff,rh} = x_{\rm pl-ff,rh,Nbody}/x_{\rm pl-ff,rh,HMC,av,scaled}$ .}} 
\hline
\end{tabular}
\label{tab:NBdata}
\end{table}

We note here that many of our encounters lead to very small changes of
the eccentricity or semi-major axis of the planetary system (see some of
the following plots). The changes are often sufficiently small as to
cast doubt on the significance of such a result. Numerical errors are
introduced in the three-body integration for HMC and by the stochastic
background noise of potential fluctuations which are present in the
direct $N$-body simulations. To distinguish broadly those encounters
which might be regarded as unreliable from the remaining results, we
define certain criteria, namely $K<80$ and $\vert\Delta\vert > 5\cdot
10^{-7}$, which help to identify the most robust results.  In some of
the following interpretative plots we show results from the full set
of encounters, and in other cases only a limited subset. The selection
of the representative data will be clearly stated in the respective
paragraphs or figure captions.

\subsection{Dissolution of planetary systems}

First, we examine the overall statistics on planetary-system
retention.  In Table~\ref{tab:HMCdata} we provide some basic data
which are generated with the Monte Carlo (HMC) scheme.  The
simulations were stopped after a few million encounters had taken
place between planets and passing stars.  The actual evolutionary
duration corresponds to some 5-10 initial half-mass crossing times.
The table summarizes some interesting information on the dissolution
of planetary systems and the creation of free floaters in the HMC
models.  For comparison, we also present in Table~\ref{tab:NBdata},
the analogous data which are generated with the $N$-body scheme.

For all models, we provide, in the tables, the rates of free floater
liberation, alternatively in terms of the cluster's crossing or
relaxation time. In the models which represent a population of soft
planets, we create about one free floater per crossing time and about
100 per relaxation time ($N$-body). In contrast to previous claims
(Smith \& Bonnell 2001), only very few planets escape during the
whole simulation - one order of magnitude less than there are free
floaters, even though the initial orbital velocity of typical planets
is larger than the velocity dispersion of the cluster.

In order to rule out the possibility that the retention of disrupted
planets in the shallow potential of the host cluster may be due to an
under-representation of close encounters by the HMC scheme, an analogous
model is simulated with the $N$-body scheme.  In order to compare the
rates of free floater and escaper creation between these two
approaches, which were obtained for practical and technical reasons
with different numbers of particles and planetary systems,
we first have to apply an appropriate scaling factor.
In our system of $N$-body units (used here for both methods) the total
mass is constant (unity) and individual stellar masses scale as $1/N$. 
Rates per crossing time scale only by the number of planets, 
because in $N$-body units the cross section, if gravitational
focusing prevails, scales inversely 
proportional to $N$ and the stellar number density is proportional to $N$. 
However, in the comparisons between
the rates per relaxation time we should additionally scale by
the ratio of half-mass relaxation times for the two models.
The exact scaling expression
is given in the caption of Table~\ref{tab:HMCdata}.

The tables show that both the rates per crossing time per planet, and
the (scaled) rates per relaxation time agree well,
but the numbers differ, in the sense that the rate of free floater creation
is larger by some 35 \% in the $N$-body system. However, the
intrinsic variation of $N$-body results for different eccentricities is
as large as that, therefore we conclude that our HMC and $N$-body results
do agree reasonably with each other.
The $N$-body result is
that approximately  100 dissolutions of planetary systems occur
per relaxation time in this case, regardless of their initial
eccentricity.  Contrary to the standard expectation that highly
eccentric planetary systems should be more prone to disruption (Hurley
\& Shara 2002), we find that the disruption rate is essentially
insensitive to the planets' original eccentricity. In fact, there is
a slight tendency for fewer disruptions, in the case where all
planetary systems are initialized with $e=0.99$. 

In order to understand this puzzling result, we have examined the
encounter activity between planetary systems and stars in the $N$-body
models.  We find that, for highly eccentric planetary systems, there
is indeed more encounter activity (e.g. more KS terminations) by some
40\% (as compared to models with modest eccentricities, e.g. 0.6).
Nevertheless, these encounters do not necessarily lead to
free-floaters. After a brief episode of temporary liberation (KS
termination), some of the highly eccentric planets quickly become
attached to intruding stars.  The exchange of the host stars occurs as
a consequence of three- or more-body interaction.  The many-body
effects may also contribute to the small residual differences in the
number of disrupting planetary systems between the HMC and $N$-body
simulations because, in general, the many-body effects should increase
the dissolution rate (as seen in the $N$-body simulations).  This
difference vanishes for the simulations of stellar encounters with
hard planetary systems because the stellar cluster and the planetary
systems are dynamically well segregated and the many-body effect
should play a lesser role.  While this effect is potentially very
interesting, it is not in the scope of this paper to study it in
further detail.

\subsection{Semi major axis and eccentricity changes}

With the aid of analytic formulae, we now analyse the consequences
of stellar encounters including the vast majority which did not lead
to the dissolution of the planetary systems.  The main objective in
this section is to determine the cross sections for eccentricity and
semi major axis changes. 

\subsubsection{Domains of encounter classes}
\begin{figure}[t!]
\begin{center}
\resizebox{0.65\hsize}{!}{\includegraphics[angle=-90]{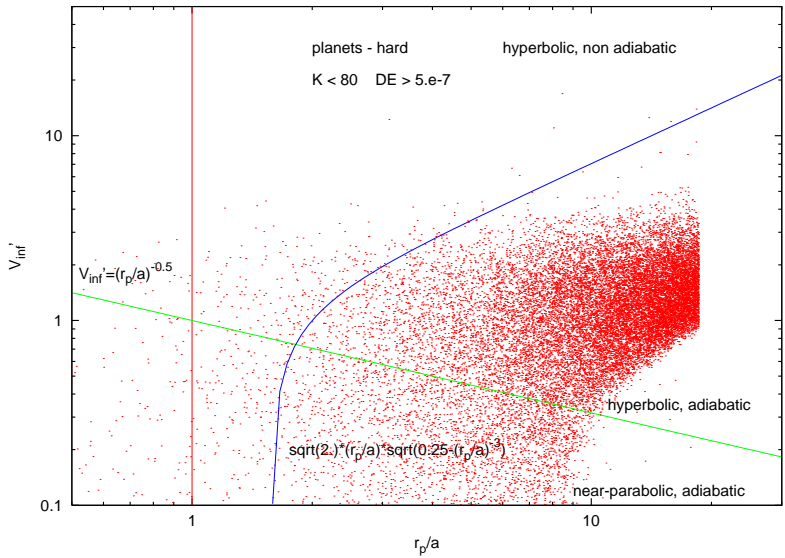}}
\caption{Location of encounters for hard planetary systems in the
hybrid Monte Carlo model, plotted according to the scaled velocity at
infinity and the minimum distance $r_p$ in units of the semi-major
axis $a$ of the planetary system. Solid lines indicate boundaries
between encounters which are close or wide, adiabatic or
non-adiabatic, hyperbolic or near-parabolic. All details of scaling
and definition of the boundaries are given in the main text: note that
they are analogous to those in Fig.~1 of Heggie 2005.}
\label{fig:scatter-hard}
\end{center}
\end{figure}

\begin{figure}[t!]
\begin{center}
\resizebox{0.65\hsize}{!}{\includegraphics[angle=-90]{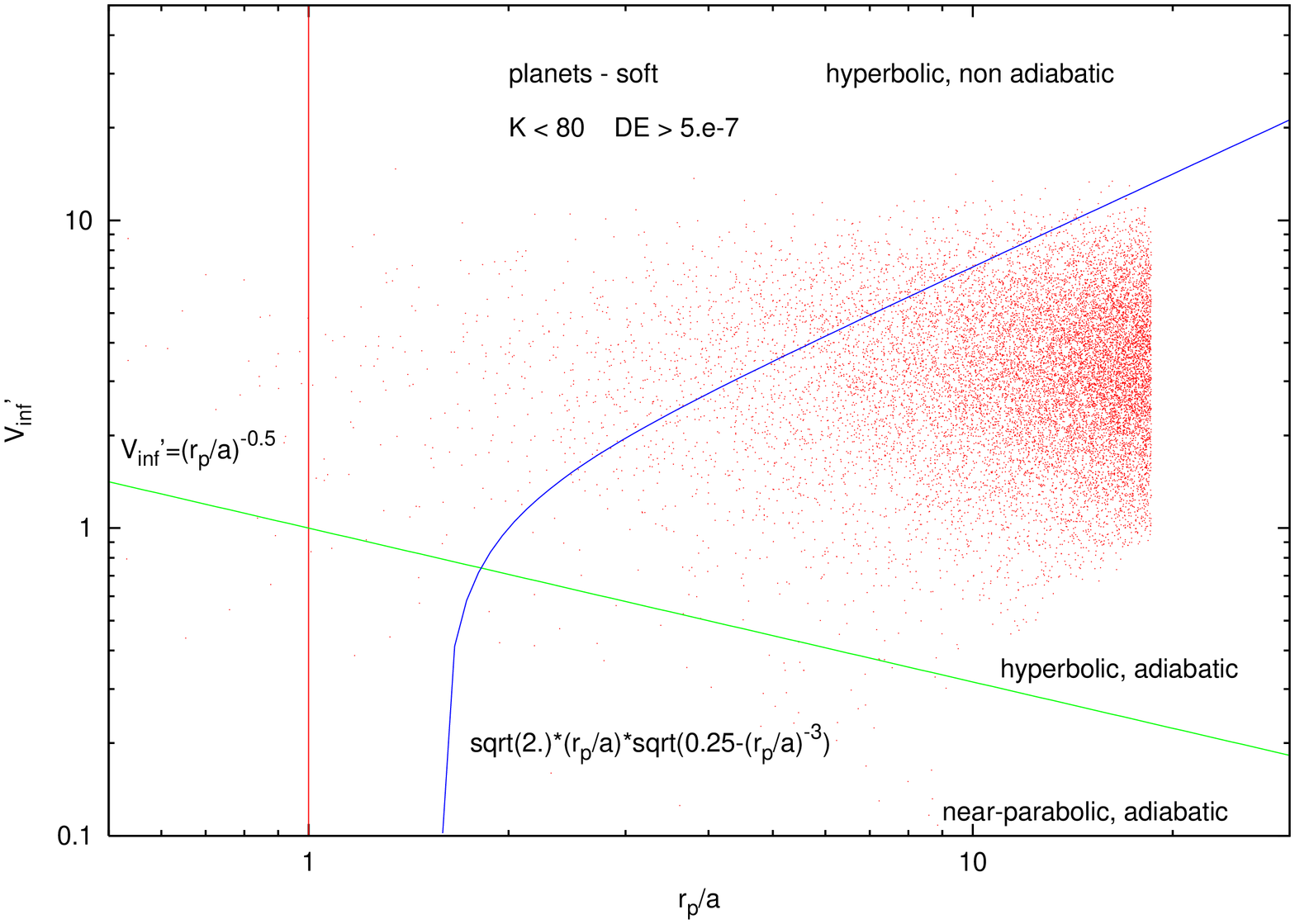}}
\caption{Same as Fig.~\ref{fig:scatter-hard}, but for soft planetary
systems.} \label{fig:scatter-soft}
\end{center}
\end{figure}

In the analysis of the numerical data, it is useful to make direct
comparison with the analytic results in \S2. Following the
prescriptions in \S2 (Heggie 2005), we first show the location of the
limited set of encounters in the normalized relative velocity-impact
parameter plane for the hard and soft planetary systems (Figs.~
\ref{fig:scatter-hard} and \ref{fig:scatter-soft}). In these figures,
the ordinate is $V/\sqrt{GM_{123}/a}$, while the abscissa is given by
$r_p/a$ for each encounter.  The parameter space is separated by three
domains in accordance with the nature of the encounters. The vertical
line ($r_p = a$) separates very close interactions from tidal
encounters. The line separating adiabatic from non-adiabatic
encounters is defined by $V/r_p = v_c/a$, where $v_c=\sqrt{GM_{12}/a}$
is the circular velocity of the planet. The line which separates
near-parabolic from hyperbolic encounters is given by the condition
$e' = 2$.

Figure.~ \ref{fig:scatter-hard} represents a model for the hard
planetary systems. Only a small number of encounters (for small
$r_p$) with hard planetary systems are non-adiabatic. There is a
considerable number of near-parabolic encounters for $r_p >a$.  In
contrast, the cloud of representative points shifts upward in the model
for the soft planetary systems (Fig.~\ref{fig:scatter-soft}).  This
systematic difference is due to the planets' larger semi-major axes.
Consequently, there are many non-adiabatic encounters with
hyperbolic speeds and $r_p > a$ for the soft planetary systems. The
number of near-parabolic encounters is negligible. Very small changes
in $\vert\Delta\vert (< 5\cdot 10^{-7}$) may be affected by 
numerical errors. The omission of 
these potentially spurious points accounts for the absence of points
in the lower right corner of the figures.  The condition that $K<80$ also
truncates the distribution of data points on the right hand side.  
A corresponding set of figures cannot be constructed for
the $N$-body simulations due to the lack of complete dynamical
information.  In these simulations the magnitude of $v$ for individual
encounters is not well determined.

\subsubsection{Correlated changes of orbital elements}
\begin{figure}[t!]
\begin{center}
\resizebox{0.65\hsize}{!}{\includegraphics[angle=-90]{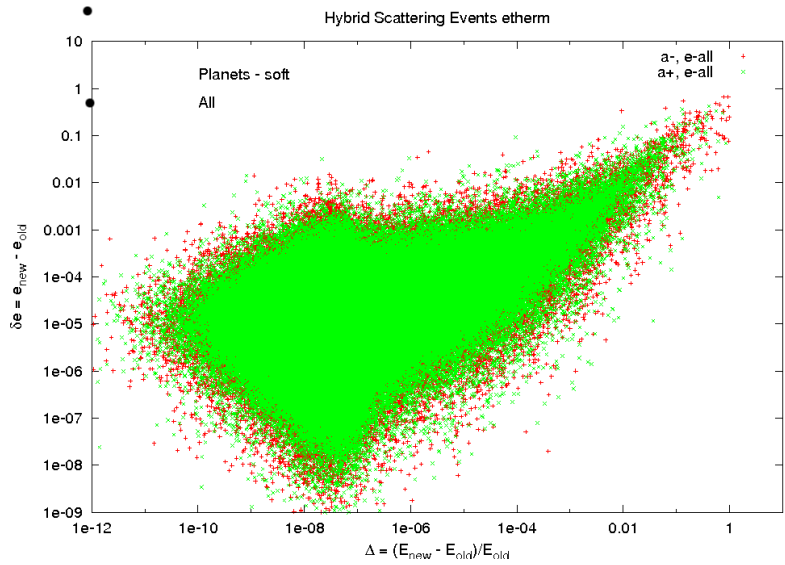}}
\caption{Relative energy change vs. eccentricity change for soft
planetary systems in the hybrid Monte Carlo model. All encounters are included.}
\label{fig:scatter-MC}
\end{center}
\end{figure}

\begin{figure}[t!]
\begin{center}
\resizebox{0.65\hsize}{!}{\includegraphics[angle=-90]{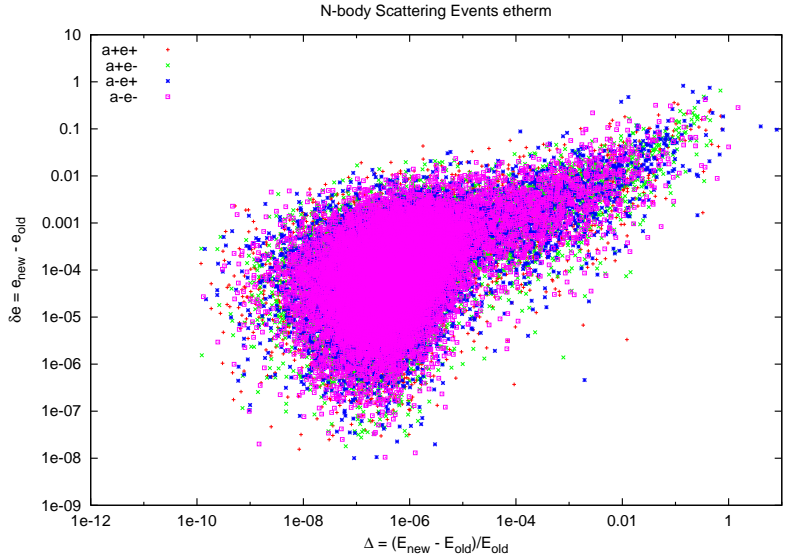}}
\caption{As Fig.~\ref{fig:scatter-MC}, but for the direct $N$-body
model (soft-planetary systems) with an initially thermal 
eccentricity distribution.}
\label{fig:scatter-NB}
\end{center}
\end{figure}

We now analyze the correlations between the relative  change in the planet's energy
and the magnitude of $\delta e$. Figures.~\ref{fig:scatter-MC} and
\ref{fig:scatter-NB} illustrate the full set of encounters for the
soft planetary systems in the {} HMC and $N$-body models (run E),
respectively. Here we can clearly identify a huge number of encounters
with very small changes. The data points on the upper right hand
corner of these figures represent planetary systems with modest
changes in both the relative energy and eccentricity.  These figures
show that there is a correlation between the relative semi-major
axis and eccentricity changes.

This correlation is much weaker for the bulk of encounters with very
small changes, which is one of the reasons we consider changes at this
level may be due to spurious random noise.  For the $N$-body models,
the uncorrelated data representing very small orbital changes may also
be due to the unresolved potential fluctuations and multiple
concurrent encounters (see discussion above on how we sample the
encounter data in the $N$-body model).  In the HMC model, where
the maximum impact parameter in the simulation was set to be 2000 AU,
even the weakest encounters are well defined.  The uncorrelated
measurement of energy and eccentricity changes represents the small
limits of numerical accuracy in the few-body integration and in
their initialisation.

As in the $N$-body data, however, for the data analysis we only use those
encounters which obey 
$\vert \Delta\vert , \vert \delta e \vert < 5 \cdot 10^{-7}$ to exclude
numerically unreliable results (see discussion in Sec. \ref{sec:nb}).
Both figures show that the
changes are very symmetric with respect to their sign, i.e. there are
as many cases with eccentricity decrease as with eccentricity
increase. Note that for the system with larger $N$ (the HMC models
have 300,000 stars, as compared to the $N$-body models, which had only
19,000), the median values of $\Delta$ and $\delta e$ are roughly one
order of magnitude smaller. This difference is consistent with the
interpretation that encounters become weaker for larger systems.

\begin{figure}[t!]
\begin{center}
\resizebox{0.65\hsize}{!}{\includegraphics[angle=-90]{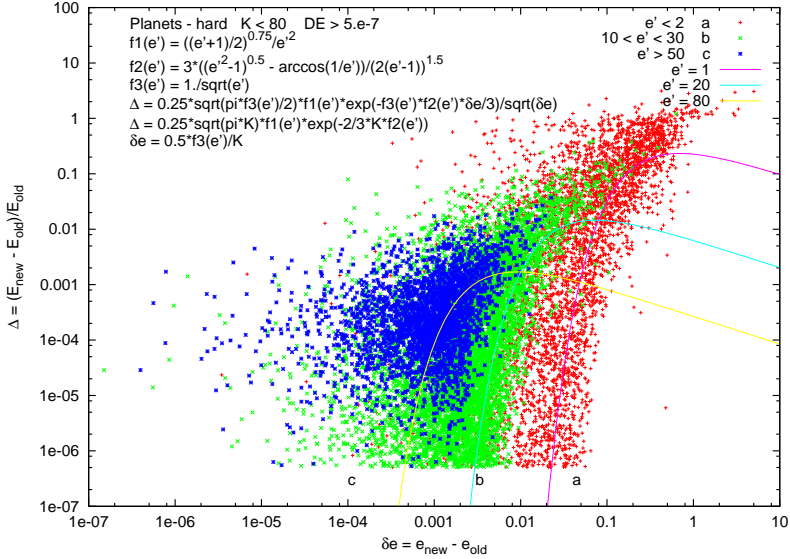}}
\caption{Location of encounters for hard planetary systems in the
hybrid Monte Carlo model, plotted by relative energy change and
eccentricity change, for three cases with different $e'$
(near-parabolic, intermediate, and extremely hyperbolic; see main text
for more details). For comparison solid lines are plotted from
analytic expressions of Heggie \& Rasio (1996) and Heggie (2005).}
\label{fig:eccen-energy-hard}
\end{center}
\end{figure}

\begin{figure}[t!]
\begin{center}
\resizebox{0.65\hsize}{!}{\includegraphics[angle=-90]{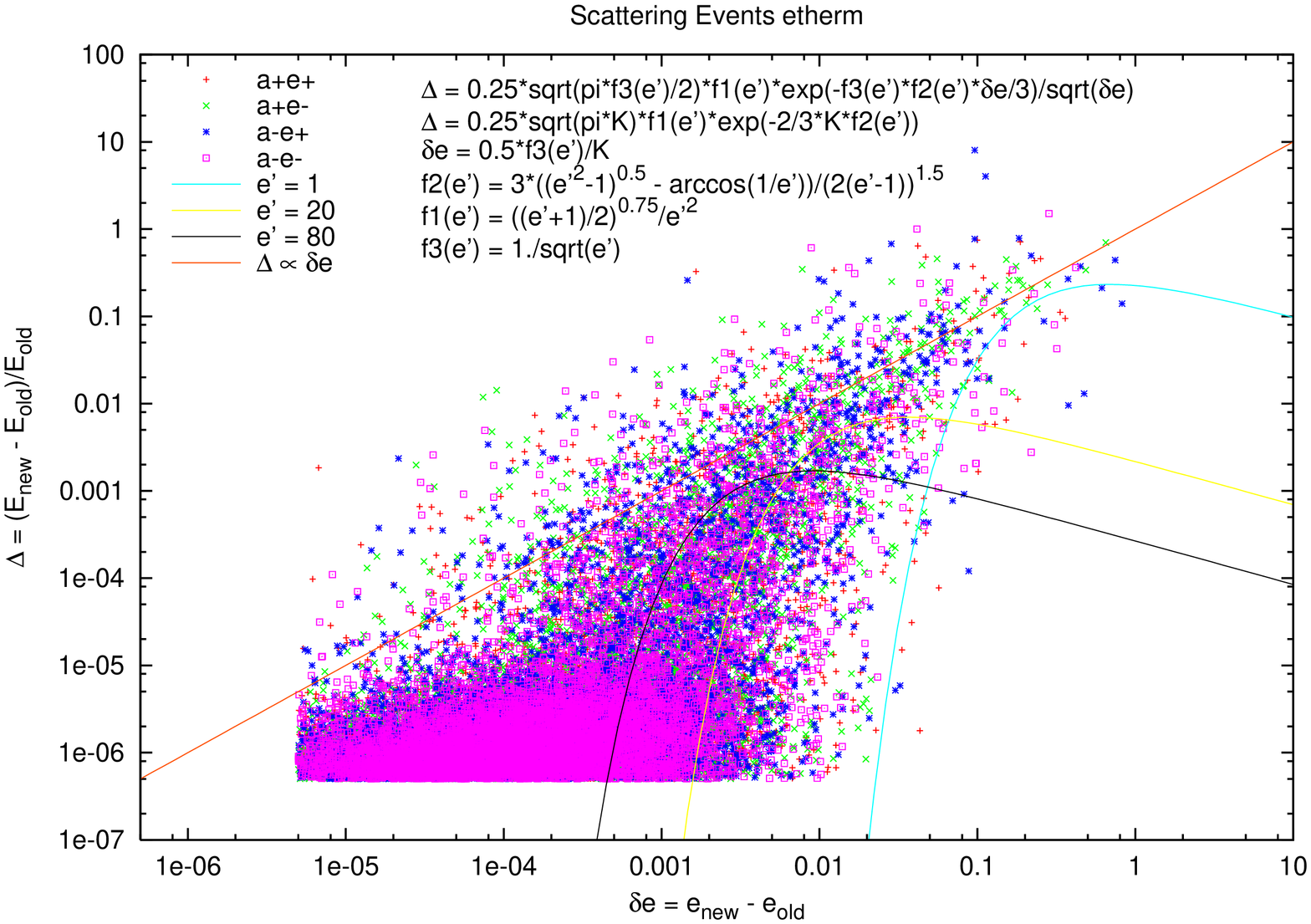}}
\caption{Location of encounters for planetary systems with an
initially thermal eccentricity distribution in the direct $N$-body
model (soft-planetary systems), plotted by relative energy 
change and eccentricity change.  For
comparison the same solid lines as in Fig.~\ref{fig:eccen-energy-hard}
are plotted, based on analytic expressions of Heggie \& Rasio (1996)
and Heggie (2005).}
\label{fig:eccen-energy-NB-th}
\end{center}
\end{figure}

\subsubsection{Parabolic versus hyperbolic encounters}

The next three Figs.~\ref{fig:eccen-energy-hard},
\ref{fig:eccen-energy-NB-th} and \ref{fig:eccen-energy-NB} again
illustrate the dynamical changes in the plane defined by $\Delta $ and
$\delta e$.  The main purpose of this analysis is to show the
dependence of the outcome on the initial relative speeds between the
encountering stars. Indeed it was the study of these results
which forced us to extend theoretical results in the literature to
the case of hyperbolic encounters.

For the HMC model (Fig.~\ref{fig:eccen-energy-hard}) we show three
sets of encounters, depicted by different colours of points: $e'<2$
(red), $10 < e'< 30$ (green), and $e' > 50$ (blue); the general limits
defined above ($K<80$, $\Delta > 5 \times 10^{-7}$) were also applied.
These three groups of data points represent nearly parabolic, intermediate,
and hyperbolic encounters.  Note that most hyperbolic encounters lead
to very small changes in $a$ and $e$. Only the parabolic
encounters satisfy the necessary condition for planetary disruption and
major orbital element changes, i.e. $\vert \Delta \vert > 1$ and
$\delta e \sim 1$ respectively.

For the $N$-body model (Figs.~\ref{fig:eccen-energy-NB-th} and
\ref{fig:eccen-energy-NB}), although the encounters cannot be
individually distinguished by their $e'$, the overall results agree
with those obtained from the HMC model. (Here data for very small
changes, i.e. with $\delta e < 5 \times 10^{-6}$, are omitted, see
figure). Figure~\ref{fig:eccen-energy-NB-th} shows $N$-body results
with an initially thermal eccentricity distribution, while
Fig.~\ref{fig:eccen-energy-NB} presents results from run 1 (with $e_0
= 0.01$ initially for all planetary systems). 

In order to compare these numerical data with the analytic results
(see \S2), we super impose in all three figures, the correlated
magnitude of $\delta e$ and $\Delta$ derived from the analytic
expressions in equations~(\ref{eq:de0}) and (\ref{eq:Delta}), for
three different values of $e' = 1, 20, 80$, respectively. The results
of the HMC simulations and those of the $N$-body model E (with an
initial thermal eccentricity distribution), i.e. 
Figs.~\ref{fig:eccen-energy-hard} and \ref{fig:eccen-energy-NB-th},
agree fairly well with each
other and with theory. Both models indicate that planets break up
primarily through parabolic encounters.  Note that $e'$ increases
from right to left in all three figures.  It can also be seen that the
encounters which have been excluded (due to small changes or large
$K$) are extremely hyperbolic.

\subsubsection{Planetary systems with nearly circular initial orbits
}\label{sec:nearly_circular}

\begin{figure}[t!]
\begin{center}
\resizebox{0.65\hsize}{!}{\includegraphics[angle=-90]{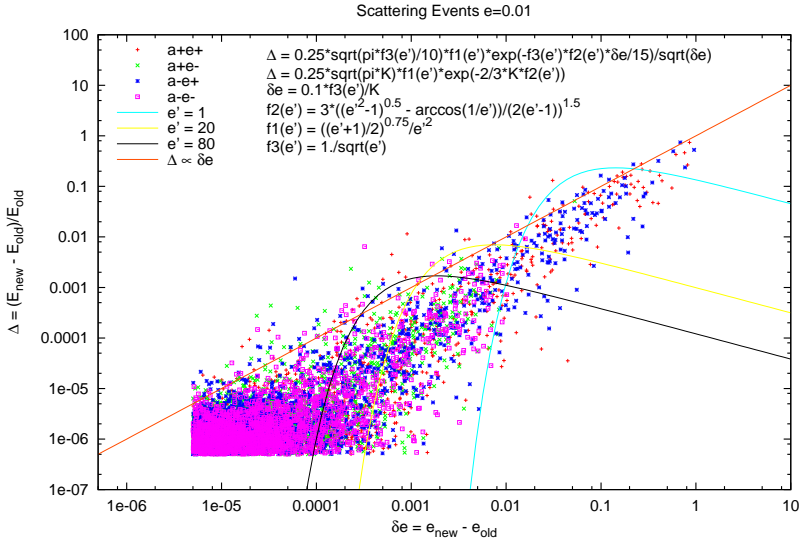}}
\caption{Location of encounters for planetary systems with initial
eccentricity $e_0 = 0.01$ in the direct $N$-body model for soft 
planetary systems, plotted by relative energy change and eccentricity 
change. Again the same solid lines as in
Fig.~\ref{fig:eccen-energy-hard} are plotted from analytic
expressions of Heggie \& Rasio (1996), Roy \& Haddow (2003) and
Heggie (2005).} 
\label{fig:eccen-energy-NB}
\end{center}
\end{figure}

This case requires separate discussion.  
The plot of the $N$-body results for $e_0=0.01$ (nearly circular
orbits, Fig.~\ref{fig:eccen-energy-NB}) shows some differences from the
previous two cases.  For planetary systems
which suffer large orbital changes, the correlation between $\delta e$
and $\Delta $ is much more pronounced, with less scattering of the
points at the upper right hand side of the figure. For a given
$\delta e$ these encounters for nearly circular binaries
tend to exhibit smaller energy changes than in the previous cases. 

Now we turn to a theoretical interpretation of these differences.
Several theoretical results
change when one considers {\sl strictly} circular initial orbits
(cf. Appendix \ref{sec:e-change}).  This can also be seen in 
Fig.~\ref{fig:eccen-energy-NB}, where we have plotted curves based on
assuming the same dependence of $\delta e$ and $\Delta$ on $e'$ as for non-circular
orbits.  This
comparison is unsatisfactory, particularly for the small values of $K$, which
correspond to the limit of strong encounters.

First consider the encounters for which $\vert\delta e\vert\gg e =
0.01$.  In this case analytical theory for a circular binary is
relevant.  In eqs.(\ref{eq:de0circ}) and (\ref{eq:de-circ-parabolic}), which give the change
in eccentricity for this case, the main dependence on the distance of
closest approach is in the exponential factors.  The same exponential
dependence occurs in the formula for the relative energy change in the
parabolic case (eq.(\ref{eq:cdeps})) and in the hyperbolic case (not
shown).  Therefore we may expect $\Delta\propto\delta e$ for large
changes in eccentricity (and energy).  The dashed line in
Fig.\ref{fig:eccen-energy-NB} has the same slope, and it is followed
by the trend of the points for large $\delta e$ and $\Delta$.

There is a further complication, because the figure we are discussing
deals with soft planetary systems, and the closest encounters are
distinctly non-adiabatic (Fig.\ref{fig:scatter-soft}), and indeed
impulsive.  To find out how $\delta e$ and $\Delta$ are correlated in
this case we may make use of eqs.(\ref{eq:dE}), (\ref{eq:dJ}), from
which it can be deduced that
$$
\delta(e^2) = (1-e^2)\left\{-\frac{\delta E}{E} -
\frac{\delta(J^2)}{J^2}\right\}.
$$
Now for an impulsive encounter with a circular binary $\delta E =
\mybv.\delta\mybv$ and $\delta J^2 = 2a^2\mybv.\delta\mybv$.  But we also have
$J^2 = GMa$ in this case, where $M$ is the total mass of the binary,
and a quick calculation shows that the lowest order contribution to
$\delta(e^2)$ vanishes.  (It is obvious that it must do, as this
contribution would be proportional to $\mybv.\delta\mybv$, and would give
both positive and negative values, whereas we must have $\delta(e^2) >
0$ for an initially circular binary.)  It follows that $\delta(e^2)$ must
be proportional to $\vert\delta\mybv\vert^2$, in this case, and so
$\delta e\propto\vert\delta\mybv\vert$.  But $\delta E$ is also
proportional to $\vert\delta\mybv\vert$, as mentioned in
Sec.\ref{sec:imp}, and so again we deduce that $\Delta\propto\delta
e$. 

In this case of
impulsive encounters, it was mentioned in Sec.\ref{sec:impulse} that
$\Delta\sim D \equiv \delta(e^2)$.   This suggests a steepening of the
trend in the dependence of $\Delta$ on $\delta e$ at the extreme right
of the diagram, and this is not observed. 

In the remainder of the diagram ($\vert\delta e\vert\ll0.01$) the
binary essentially behaves as one with a non-zero initial
eccentricity, and the distribution of points may be expected to behave
much as in the two previous cases.  The analytical expressions for
$\delta e$ and $\Delta$, i.e. eqs.(\ref{eq:de-noncirc}) and
(\ref{eq:deps}), are approximately proportional to $e$ for small $e$,
and so the curves in this diagram should be scaled by a factor of
order $0.01$.  This brings them into better accord with the points
corresponding to more distant encounters.  The numerical results indicate 
that the amplitude of eccentricity change is considerably larger than 
that of the change in energy or semi major axis.  
Since the dissolution rate is essentially
independent of the initial eccentricity, planets which formed with
circular orbits can quickly become eccentric.

\begin{figure}[t!]
\begin{center}
\resizebox{0.65\hsize}{!}{\includegraphics[angle=-90]{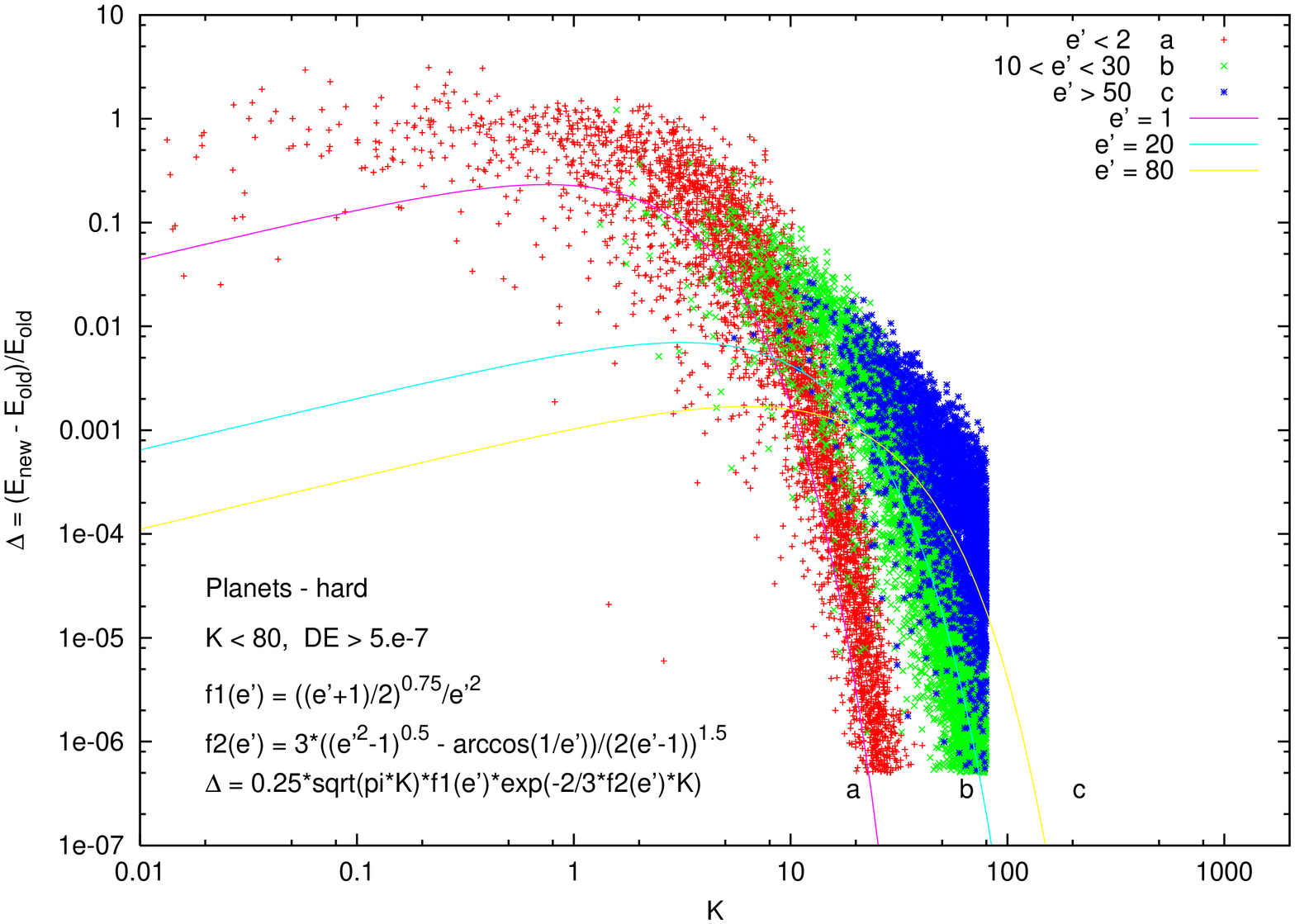}}
\caption{Location of encounters for hard planetary systems in the hybrid
Monte Carlo model, plotted by relative energy change as a function
of $K=(r_p/a)^{3/2}$, for three cases with different $e'$
(near-parabolic, intermediate, extremely hyperbolic; see main text for more
details). For comparison solid lines  based on
analytic expressions of Heggie (2005) are plotted.} \label{fig:K-energy-hard}
\end{center}
\end{figure}

\subsubsection{Dependence on the intruding star's distance of closest 
approach}

In \S2, we show that the magnitude of a planet's dynamical response to
a stellar encounter depends on both the impact speed (through $e'$) and
the distance  of closest approach (through $K$).  We now interpret our
numerical results in terms of the dependence of $K$ given by
the analytic expressions. In Figure~\ref{fig:K-energy-hard} the dynamical changes
in the planetary orbits are depicted as in the three figures before,
but this time plotted in the plane of $K$ and $\Delta$. Only HMC 
simulations are represented, for $N$-body models it was impossible to obtain
reliable encounter parameters for each interaction and compute $K$ 
(see Sec.~\ref{sec:nb}). This plot can be compared with the
analytic results more directly because the latter are often expressed as
functions of $K$, whereas the correlation with $\delta e$ is more
indirect. 

The numerical results obtained with the HMC scheme (Fig.~\ref{fig:K-energy-hard})
agree very well with the analytic expression. In this case, the fractional energy
change at large $K$ is correlated with  $e'$.  This trend
indicates that in the high $K$ limit, a vast majority of encounters
are very hyperbolic.  Only nearly parabolic ($e' < 2$) and non adiabatic,
close (small $K$) stellar encounters lead to planets' disruption, i.e. $\vert
\Delta \vert >1$.  In addition, it follows from
Fig.~\ref{fig:eccen-energy-hard} that most non adiabatic, close
encounters lead to major eccentricity changes, i.e. $\vert \delta e \vert
\sim 1$.

\begin{figure}[t!]
\begin{center}
\resizebox{0.65\hsize}{!}{\includegraphics[angle=-90]{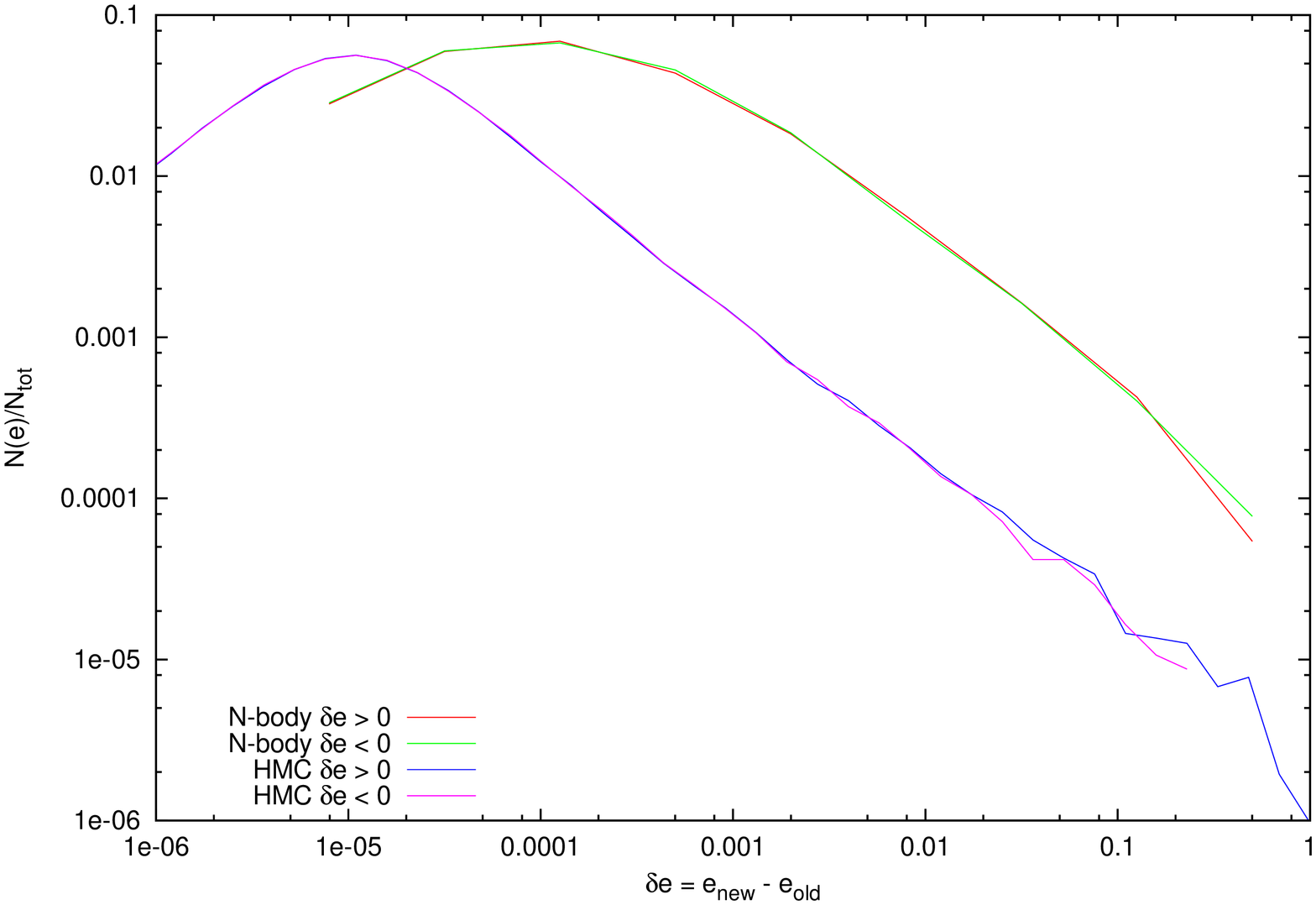}}
\caption{Fractional number of encounters as a function of
eccentricity change (soft
planetary systems), compared between the hybrid Monte Carlo  and
$N$-body models with an initially thermal
eccentricity distribution. The data are given separately for
positive and negative changes.} 
\label{fig:HM-de}
\end{center}
\end{figure}

\begin{figure}[t!]
\begin{center}
\resizebox{0.65\hsize}{!}{\includegraphics[angle=-90]{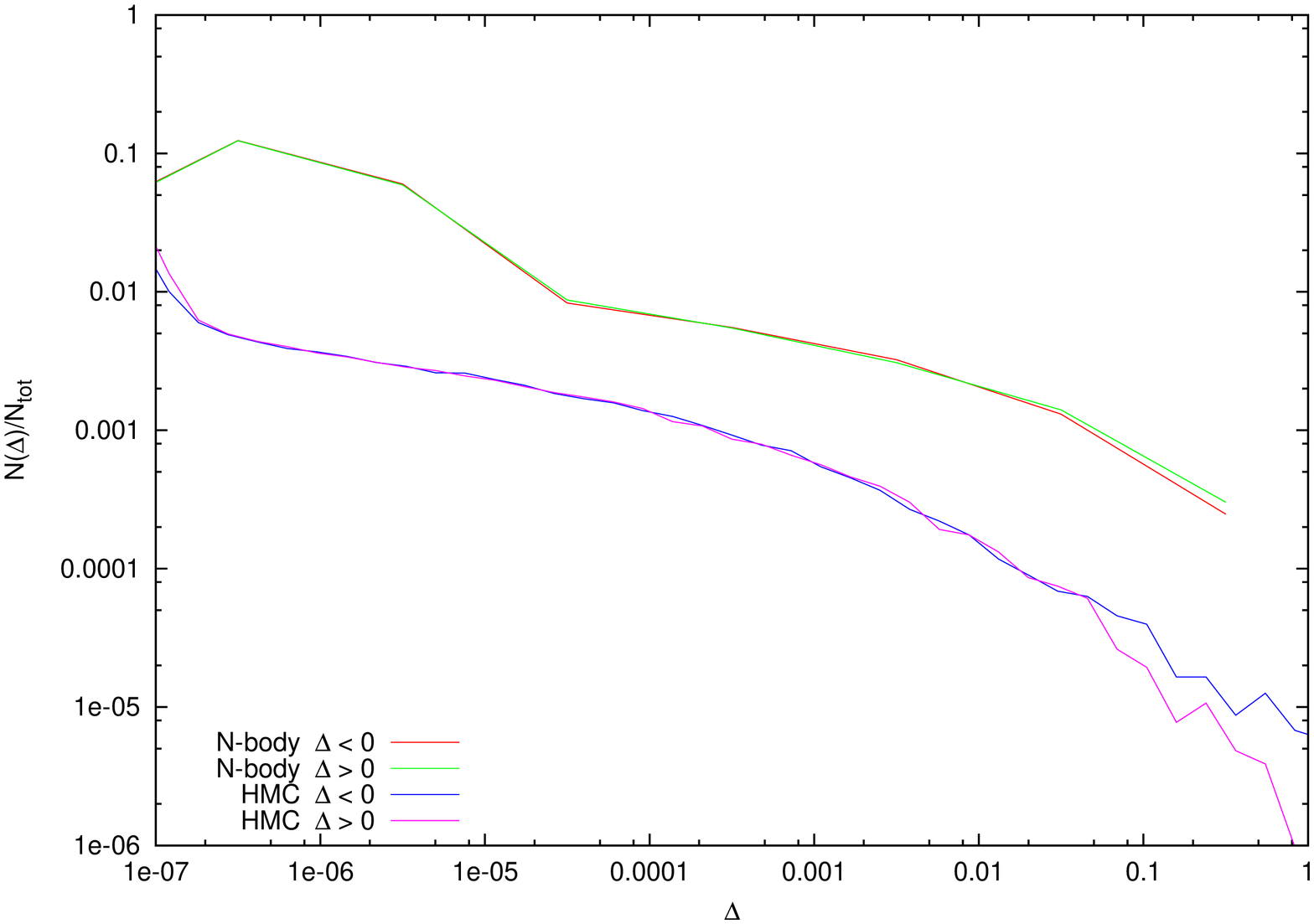}}
\caption{Fractional number of encounters as a function of relative
energy change (soft planetary systems), compared between the
hybrid Monte Carlo and $N$-body models with an initially thermal
eccentricity distribution. The data are given separately for positive
and negative changes. } \label{fig:HM-da}
\end{center}
\end{figure}

\subsection{Rates of energy and eccentricity evolution}

\subsubsection{Distributions of $\delta e$ and $\Delta$}

For another comparison of the HMC and NBODY6++ results we have just
binned the changes of $\Delta$ and $\delta e$ and computed the numbers
$N(\Delta)$, $N(\delta e)$ normalized to the total number of
events. Figs.~\ref{fig:HM-de} and \ref{fig:HM-da} show the results as
a function of $\delta e$ and $\Delta$, respectively. Note that this is
an unnormalized measure of the total cross sections. 
In the following section we will deduce properly
the normalized differential cross sections as well. 

Here, we first observe that the shapes of the encounter frequency
distributions are similar, albeit they are shifted (the numbers for
the HMC
model are smaller by about one order of magnitude, except for small
$\vert\delta e\vert$). There are two
possible explanations for this shift, and both effects may
contribute. With a larger particle number in the HMC model all changes
are on average about an order of magnitude smaller than in the $N$-body
model (compare discussion of Fig.~\ref{fig:scatter-MC}) and so the
curves are shifted to the left. On the other hand, in the $N$-body
system individual encounters with extremely small changes (which are
easily observed in the HMC model) are obscured by the background of
stochastic potential fluctuations. 

The nearly logarithmic distribution $N(\delta e) \propto \delta e^{-1}$ 
for $\delta e > 10^{-4}$ implies that the mean square changes of eccentricity
are dominated by close encounters and large $\delta e$.
We also note here that, in the HMC
models we observe that the initially thermal eccentricity distribution
is preserved, as would be expected.

Turning now to the distribution of $\Delta$,  we note that, for energy 
changes with $\Delta \age 0.1$,
negative changes of $\Delta $ are more probable, i.e.  that there
is a preferred trend towards softening of planetary orbits. 
This confirms the discovery by \citet{Fr06} that there is a range of intermediate
 planetary orbits with $v_c < V < v_{\rm orb}$ for which there is a net surplus of
 softening encounters as compared to hardening ones. We also confirm this with our
 analytical results, as discussed qualitatively at the end of Section 2.3.1, and
 the result is also visible in Fig.~13 (differential cross sections) with better
 statistical quality.


\begin{figure}[t!]
\begin{center}
\resizebox{0.65\hsize}{!}{\includegraphics[angle=-90]{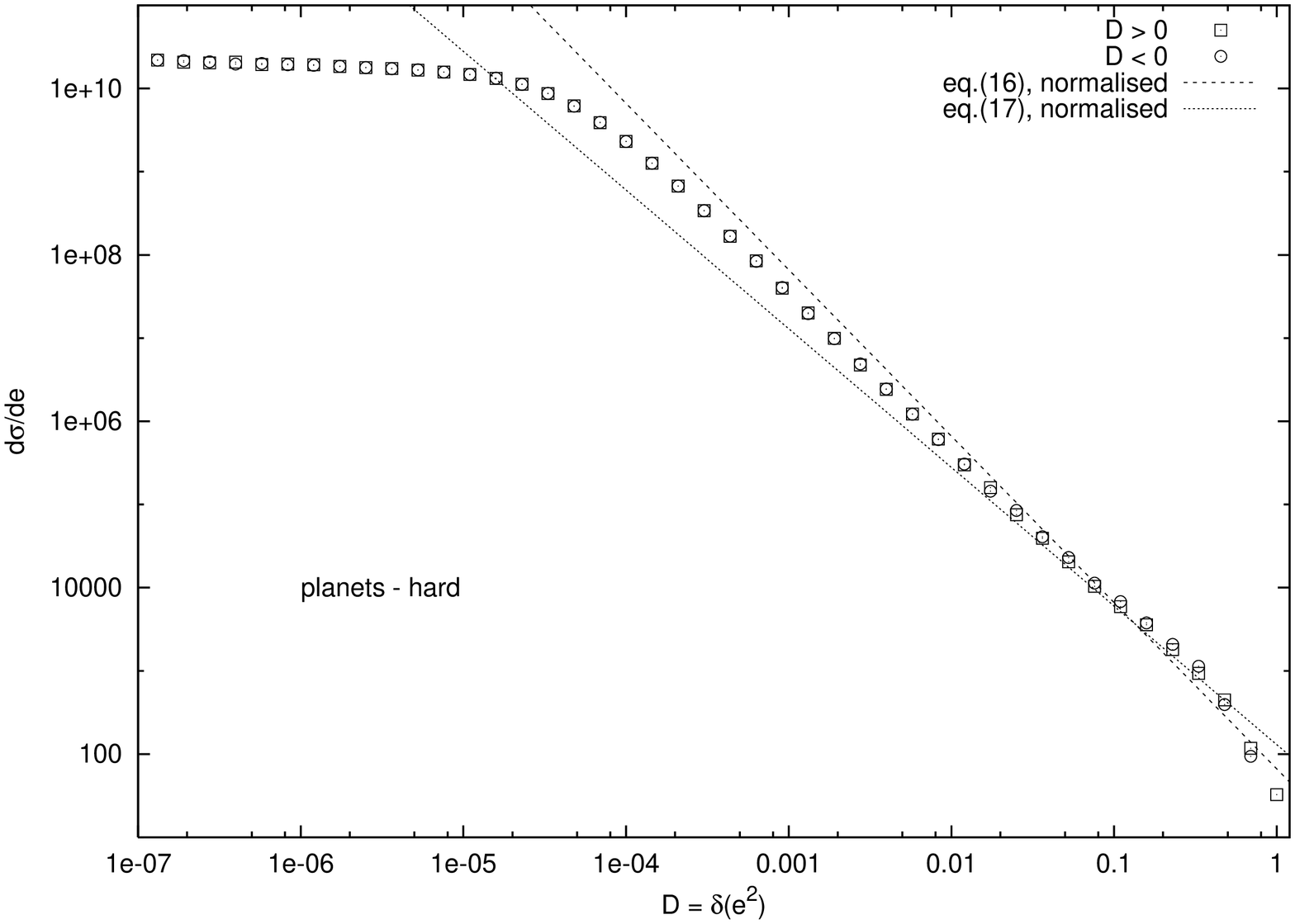}}
\caption{Differential cross section for eccentricity changes of hard
planets, as a function of eccentricity change (strictly, $D =
\delta(e^2)$) for the hybrid Monte Carlo model. The analytic results
obtained by formulae from Section~2 are plotted as straight lines,
while the binned numerical results appear as points; see detailed
explanation in the main text.}
\label{fig:eccen-hard}
\end{center}
\end{figure}

\begin{figure}[t!]
\begin{center}
\resizebox{0.65\hsize}{!}{\includegraphics[angle=-90]{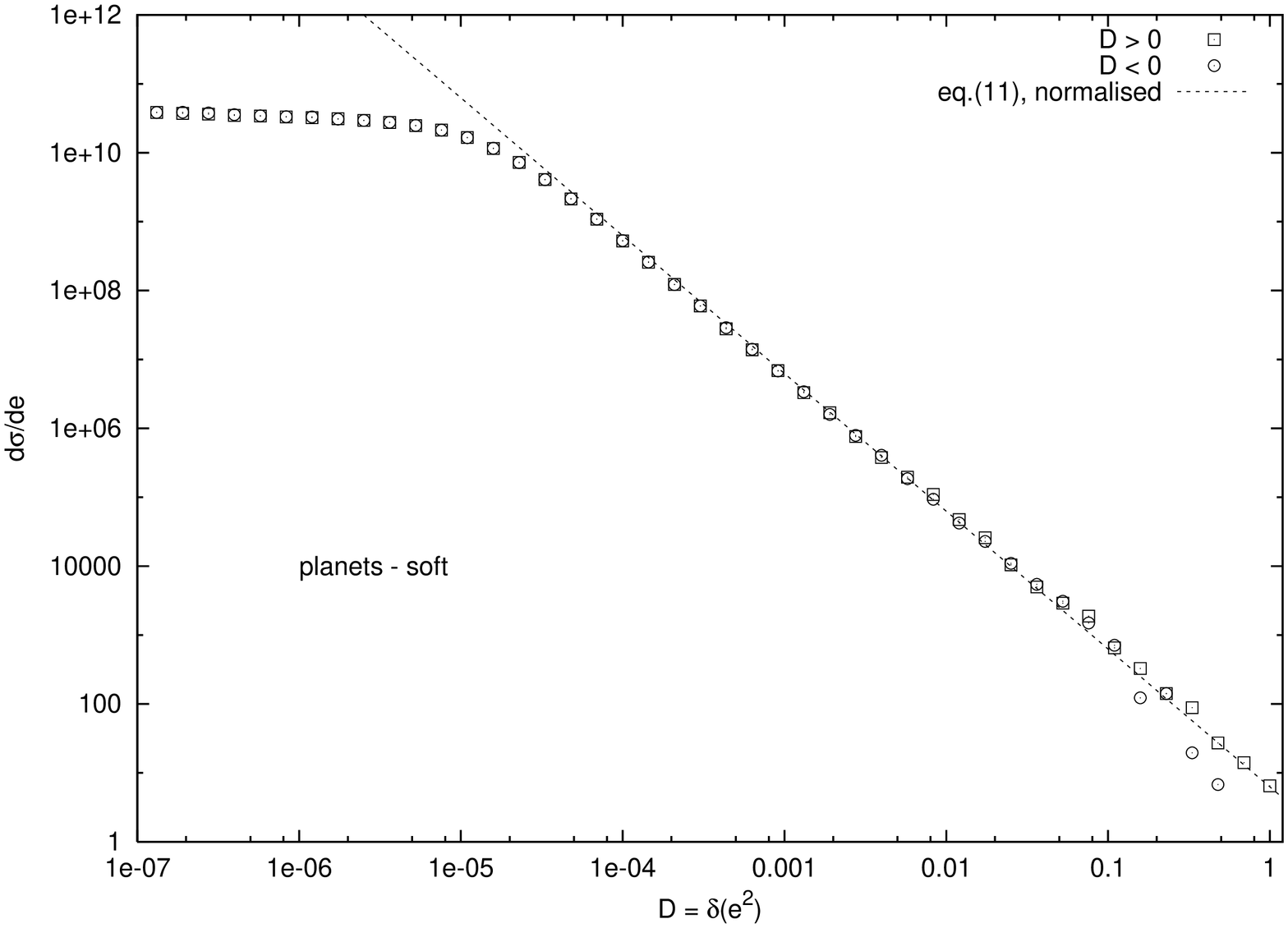}}
\caption{Same as Fig.~\ref{fig:eccen-hard}, but for soft planetary
systems.}
\label{fig:eccen-soft}
\end{center}
\end{figure}

\begin{figure}[t!]
\begin{center}
\resizebox{0.65\hsize}{!}{\includegraphics[angle=-90]{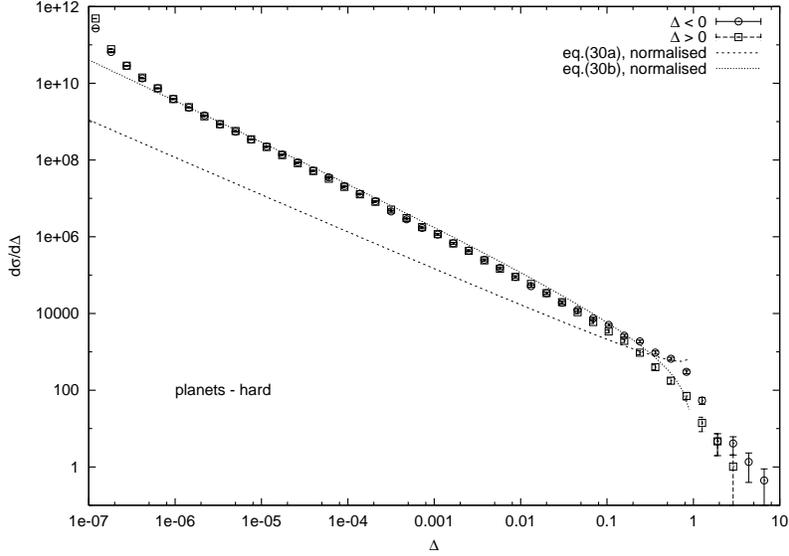}}
\caption{Scaled differential cross section for relative energy changes
of hard planets, as a function of relative energy change for hybrid
Monte Carlo model. The analytic results obtained by formulae from
Section~2 are plotted as lines, while the binned numerical results
appear as points, see detailed explanation in the main text.}
\label{fig:energy-hard}
\end{center}
\end{figure}

\begin{figure}[t!]
\begin{center}
\resizebox{0.65\hsize}{!}{\includegraphics[angle=-90]{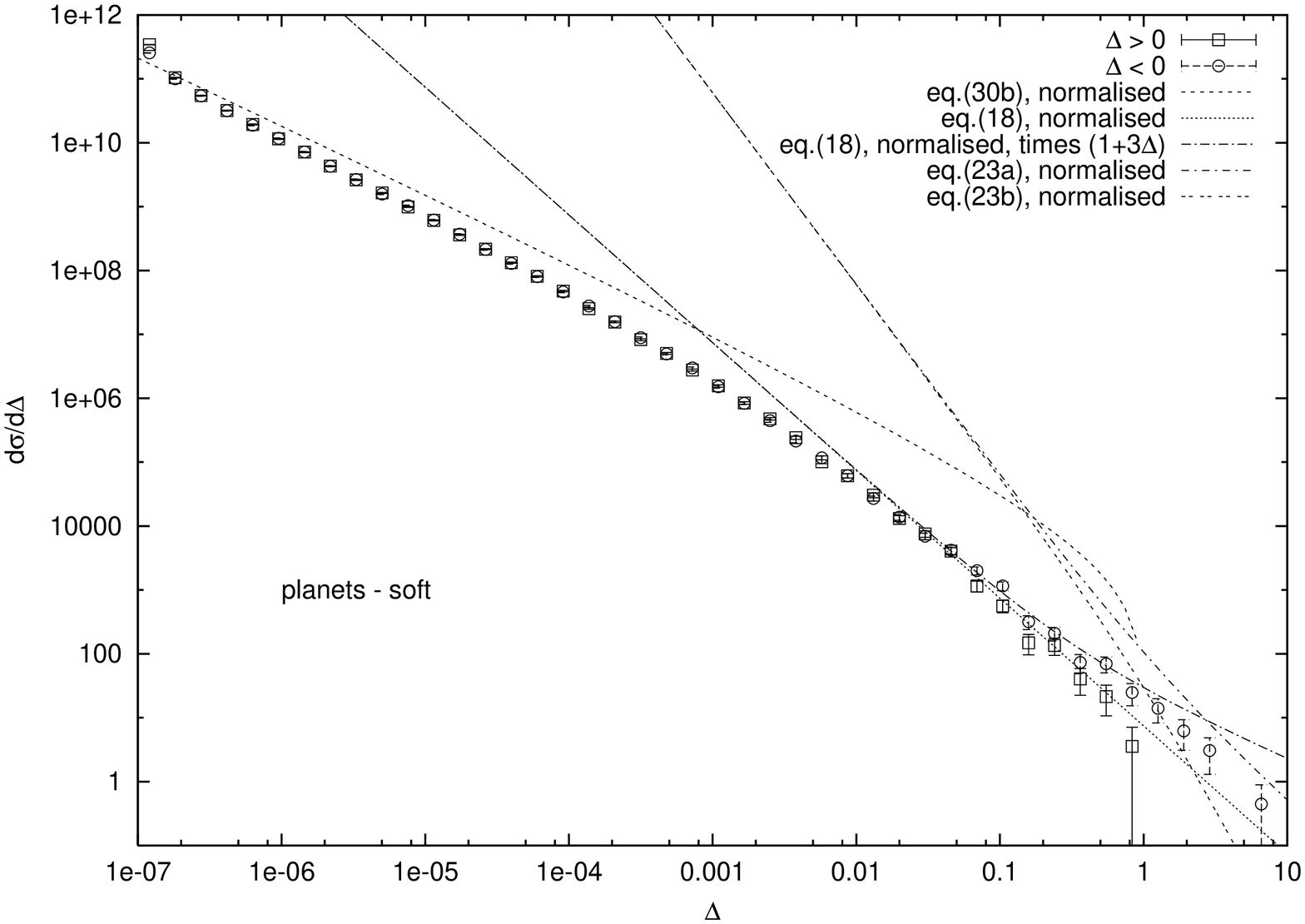}}
\caption{Same as Fig.~\ref{fig:energy-hard}, but for soft planetary
systems. } \label{fig:energy-soft}
\end{center}
\end{figure}

\subsubsection{Normalization for differential cross sections}
Finally we compute properly normalized differential cross sections,
and compare them with theory.  We restrict the numerical data to those
from the HMC runs.  Differential cross sections can be obtained from
our numerical results using the binned data $N(\Delta)$ described
previously.  Because the cross sections depend on $a$ and $V$, however
(Section~\ref{sec:analytic_results}), some post-processing is
necessary.

We define $P$ to be the probability that a given encounter results in
a value of $\Delta$ which lies within a certain bin.  Then
\begin{equation}
  P = \frac{1}{\pi p_{max}^2}\frac{d\sigma}{d\Delta}(\Delta,a,V)\delta\Delta,
\end{equation}
where $\delta\Delta$ is the range of the bin.  Let us suppose that the
differential cross section has a power-law dependence on $a$ and $V$,
i.e.
\begin{equation}
  \frac{d\sigma}{d\Delta}(\Delta,a,V) =
  \frac{d\sigma}{d\Delta}(\Delta,1,1)a^\alpha V^\beta,
\end{equation}
where $\alpha,\beta$ are certain constants and $d\sigma/{d\Delta}
(\Delta,1,1)$ is the normalization factor.  Then, equating $N(\Delta)$
to its expected value, we have
\begin{equation}\label{eq:normalisation}
\frac{N(\Delta)}{\delta\Delta} =
\frac{d\sigma}{d\Delta}(\Delta,1,1)~\sum_{\rm encounters}\frac{a^\alpha
V^\beta}{\pi p_{max}^2}.
\end{equation}
In the following four figures, the theoretical cross sections from
Section~\ref{sec:analytic_results} are scaled with the summation
factor on the right side of this equation, and plotted, along with the
numerical data, against $\Delta$ (or $D = \delta(e^2)$).  Note that
the summation factor has to be computed separately for each
theoretical expression.

Exactly the same prescription can be applied to determine the cross
section for eccentricity changes.  The resulting differential cross
sections are given in Figs.~\ref{fig:eccen-hard} and
\ref{fig:eccen-soft} for hard and soft planets, respectively.  For
comparison with analytic expressions, we also plot in these figures a
number of theoretical results from Section~\ref{sec:sigma-D},
normalized as in equation~(\ref{eq:normalisation}).

\subsubsection{Differential cross section for eccentricity evolution}

We first consider the case of hard planets (Fig.\ref{fig:eccen-hard}).
The two theoretical cross
sections in the figures correspond to the limits of parabolic (dotted)
and extremely hyperbolic (dashed) encounters.  The former correspond
to closer encounters and therefore larger values of $\vert D\vert$.
The crossover between the two formulae should occur at
$D\simeq\displaystyle{0.3\left(\frac{V^2a}{Gm}\right)^{3/2}}$.  The
trend of points with $\vert D\vert \age 10^{-4}$ indeed steepens
appropriately as $\modD$ decreases below about $0.01$. However, since
there is a considerable range in the semi-major axis, this figure
merely indicates that the range over which the change of slope takes
place is reasonable.  Figure~\ref{fig:scatter-hard} shows that there
is a large overlap (in $r_p/a$) of parabolic and hyperbolic
encounters, and so the position of the data points between the two
theoretical lines is acceptable. The notable exception is the strong
flattening below $\modD\simeq10^{-4}$. This saturation results,
however, from the chosen value of $p_{max}$.

Next, we turn to the case of soft planets (Fig.~\ref{fig:eccen-soft}),
where the interpretation, in terms of tidal impulsive encounters, is
more straightforward.  The plotted analytical result
(equation~(\ref{eq:impulsive-D})) is satisfactory down to values below
$D\sim10^{-4}$.  The saturation value in the small $D$ limit is
entirely consistent with that expected, using
equation~(\ref{D-estimate}), for encounters at a distance $r_p\sim
p_{\max}$.

\subsubsection{Differential cross section for fractional energy change}
Differential cross sections for the relative energy changes are
displayed in Figs.~\ref{fig:energy-hard} and \ref{fig:energy-soft}.
Here it is the case of hard planetary systems which is most
straightforward.  Almost all of the data agrees very well with the
analytic expression for hyperbolic adiabatic encounters
(equation~(\ref{eq:crossdelta_full})).  The apparent cutoff at
$\modDelta = 1$ is spurious, since the validity of the formula is
restricted to large values of $\vert\ln\modDelta\vert$ (or $\vert
\Delta \vert < < 1$.  Figure~\ref{fig:scatter-hard} indicates that the
closest encounters should be nearly parabolic, and so we have also
plotted the corresponding theoretical result (dashed in
Fig.~\ref{fig:energy-hard}).  It is not clear whether it plays a
significant role here.  We do notice, however, that softening
encounters (with $\Delta<0$) have a higher cross section than
hardening encounters, in accordance with the qualitative discussion at
the end of Section~\ref{sec:impulse}.

We attribute the rise in the numerical cross section at small
$\modDelta$ in Figure~\ref{fig:energy-hard} to numerical errors.  For
encounters with $p\ale p_{max}$, we estimate from
equations~(\ref{eq:Delta_hyerbolic}) and (\ref{eq:eprime}) that
$\Delta\sim10^{-44}$ (where even the exponent is an order-of-magnitude
estimate!). This magnitude is much smaller than numerical errors, and
even much smaller than the start-up error of initiating the three-body
integration at a finite distance of order $p_{max}$.  Such encounters
thus have errors which place them at the left hand side of this
figure.

The interpretation of the cross section for energy changes in soft
systems (Fig.~\ref{fig:energy-soft}) is the most complicated of all.
It is also most relevant since dissolution occurs more frequently in
this limit.  At the extreme left of this figure, the result of
numerical errors is again visible.  Over a range of small values of
$\modDelta$, the numerical results lie just below the theoretical
curve for adiabatic hyperbolic encounters
(equation~(\ref{eq:crossdelta_full})).  This $\modDelta$ dependence
intersects a steeper curve which represents the impulsive tidal
encounters (equation~(\ref{eq:Delta_impulsive})).  The numerical
results follow the steepening trend.  By comparing the two equations
we expect the crossover to occur at
$\vert\Delta\ln\modDelta\vert\simeq \displaystyle{\frac{2\pi}{3}
\left(\frac{Gm}{aV^2}\right)^{3/2}}$.  Though most of the encounters
with ``soft'' planets give values of $V^2a/(Gm)$ above 1, the median
value is only of order 3 (see Fig.~\ref{fig:scatter-soft}), and so the
crossover is expected to occur at values of $\modDelta$ no less than
about 0.1.  It seems likely that the interpretation of the data is
complicated by a considerable admixture of encounters which really
should be classified as hard.

For relatively large values of $\modDelta\age0.1$, we differentiate
positive and negative energy changes (see
equation~(\ref{eq:difference})) by multiplying the value given in
equation~(\ref{eq:Delta_impulsive}) with a factor $(1+3\modDelta)$
(Fig.~\ref{fig:energy-soft}).  In magnitude this corresponds quite
well to the separation in the data points between those for $\Delta>0$
(circles) from those for $\Delta<0$ (squares).

Also shown for soft planets are the two differential cross sections
for impulsive non-tidal encounters
(equations~(\ref{eq:Delta_non-tidal})).  Though they again display the
difference between positive and negative values of $\Delta$, it is not
clear what quantitative role they play in relation to this data.  We
presume that this is due to the fact that the systems are not really
very soft.  By comparing equations~(\ref{eq:Delta_non-tidal}) and
(\ref{eq:Delta_impulsive}), we estimate that the crossover should
occur at $\Delta\simeq\displaystyle{\frac{16\sqrt{Gm}}{V\sqrt{a}}}$.
It is only for considerably softer systems that one would expect to
see this crossover.

\section{Summary and Discussion}

\subsection{Methodological Aspects}

Very few planets are found in old open and globular clusters; we have
explored the scenario that this deficiency is associated with
disruption of planetary systems in long-lived star clusters due to
stellar encounters. Freely floating planets observed in star forming
regions support our scenario. We have studied two prototype systems
with direct $N$-body (NBODY) and hybrid Monte Carlo (HMC) simulations
of star clusters which contain a large number of planetary
systems. The direct $N$-body model represents a cluster similar to the
Hyades with 20,000 stars and 1000 planets, while the HMC models represent
a centrally concentrated globular cluster such as 47 Tuc with 300,000
stars and 30,000 planets. We have recorded changes of planetary orbits
due to encounters and the creation of free floaters (remaining in the
cluster) and escapers. Our main results are summarized in
Tables~\ref{tab:HMCdata},~\ref{tab:NBdata}.  We have shown that there
is a
consistency between analytic and numerical models in terms of the
cross section of encounters between planetary systems and a single
star, and, to first order, the two models give consistent results 
for the liberation rates of planets (creation
of free floaters).  Encounters have been measured and recorded in
large numbers and compared with analytic estimates for the changes of
semi-major axis and eccentricity. It turns out that the majority of
encounters in our simulations are well approximated by the analytic
results, which corresponds to the fact that the encounters are either
adiabatic (hyperbolic or parabolic) or impulsive and tidal in
nature. In the adiabatic regimes, evolution of the planetary
orbital elements is a diffusive process and proceeds in both
directions, i.e. with similar probabilities in eccentricity
excitation and damping as well as semi major axis increases and
declines. However, there is also a non-negligible number of encounters
which are not adiabatic and lead to relatively small minimum
distances. These close encounters give stronger changes in the orbital
parameters of planetary systems, and in particular a net injection of
orbital energy and increase in the semi major axis.

Two fundamentally different numerical methods are used in this
series of investigations.  They are the direct $N$-body and Hybrid
Monte Carlo algorithms, the latter being based on the Fokker-Planck
equation for the single stars. Both schemes provide consistent results
with each other and the analytic results (see figures and Tables). Remaining
differences e.g. in the rates of liberation of planets per crossing
time, are attributable to the more complex dynamical nature of the
$N$-body system - for many encounters with larger impact parameters
the assumption that they are uncorrelated isolated two-body encounters
is no longer fully justified. The examination of how well diffusion
coefficients for individual two-body encounters are reproduced in
direct $N$-body systems is, however, beyond the scope of this paper.
Interested readers are referred to discussions elsewhere
\citep{theuns96}.

\subsection{Application to real clusters or associations}

In order to allow the reader to extract more specific information
from our general results, we are presenting some data regarding
the free floaters in our simulations, and a scaling procedure to
estimate the corresponding results for some real clusters.

In the following Table~\ref{tab:ffdat} we provide the timescale
$\tau_{\rm ff}$ for a planet to
become a free floater, by using data of Table~\ref{tab:HMCdata},
averaged between the two physically equivalent runs).
We obtain $\tau_{\rm ff}$ by the definition 
$\tau_{\rm ff} = \tau_{\rm cr}/x_{\rm pl-Ff,cr}$
(or $\tau_{\rm ff} = \tau_{\rm rh}/x_{\rm pl-ff,rh}$, which should
give an equivalent result except for roundoff errors).
Table~\ref{tab:ffdat} provides exemplary values for 
$\tau_{\rm ff}/\tau_{\rm cr}$ and $\tau_{\rm ff}/\tau_{\rm rh}$,
respectively, which in fact are just the inverse averaged 
$x$-values taken from Table~\ref{tab:HMCdata}. 

To determine values of $\tau_{\rm ff}$ in physical units we
must employ the scaling of $\tau_{\rm ff}$ with cluster parameters.
If we assume planetary semi-major axes to stay constant in model
units, as was done in the scaling between HMC and $N$-body models
here, $\tau_{\rm ff}$ is independent of $N$, just as $x_{\rm pl-ff,cr}$.
So if a typical half-mass crossing time can be deduced from
observational data for some real clusters (as we will do below for
the Orion Nebula and the Hyades clusters)
we can just directly compute $\tau_{\rm ff}$ from that and our
measured value of $x_{\rm pl-ff,cr}$. If, however, a value of the
half-mass relaxation time is more well known from observations (as in the
case of the globular cluster 47 Tuc), we have to
be aware that $x_{\rm pl-ff,rh}$ scales with the relaxation time
(see explanations after Table~\ref{tab:HMCdata}).

\begin{table}[t!]
\caption{Time Scale $\tau_{\rm ff}$ 
for the liberation of free floaters
relative to $\tau_{\rm cr}$ and $\tau_{\rm rh}$}
\begin{tabular}{l|rrrrr} \hline\hline
        \quad Model \qquad \qquad & $\tau_{\rm ff}/\tau_{\rm cr}$      
                                  & $\tau_{\rm ff}/\tau_{\rm rh}$  
                                  & $\tau_{\rm ff}$ Gyrs (Hya)     
                                  & $\tau_{\rm ff}$ Gyrs (ONC)       
                                  & $\tau_{\rm ff}$ Gyrs (47 Tuc)  \\
HMC soft 1 & 1128.0     &  1.239    & 15.79  & 0.367     & 0.92      \\
HMC hard 1 & 54054.     &  59.52    & 756.8 & 17.57    & 44.27    \\
\end{tabular}
\label{tab:ffdat}
\end{table}

The data to obtain lifetimes of planetary systems in physical
units are obtained as follows:
The globular cluster 47 Tuc is a typical old stellar cluster with a very
high central density (of order $10^6 M_\odot/{\rm pc}^3$).  We expect
this cluster to be the most hostile environment for planetary systems.
From the Harris catalog \citep{Ha1996} we get the following data for 47 Tuc: $r_h =
2.79' $, with a distance of $R = 4.5$ kpc, which implies that $r_h=$3.6 pc; the
half-mass relaxation time is given as $\log t_{\rm rh} = 9.48$, i.e.
$t_{\rm rh}= 3.02 $ Gyrs.  In order to apply our model, we need to
determine the particle number, or total mass, which are poorly
constrained in the literature. Therefore we use the following
procedure, which is sufficient for an order-of-magnitude approach.  We
adopt the relaxation time scaling of $t_{\rm rh} \propto N
r_h^{3/2}/\log(\gamma N)$. From our HMC model, with $N=300.000$, $r_h
= 0.77 $pc, and $t_{\rm rh} = 7.28 \cdot 10^7$ yrs, we then deduce a
particle number of $N = 1.4 \cdot 10^6$ stars for 47 Tuc.  The scaling factor for
the relaxation time between our HMC model cluster and 47 Tuc is 
4.06. Hence we predict  a time scale $\tau_{\rm ff}$ to be
0.305 (14.66) relaxation times for a planet to become a free floater,
i.e. 0.92 (44.27) Gyrs.
These results have been obtained from our finding that $\tau_{\rm ff}/t_{\rm rh}$
scales inversely with the relaxation time, because in systems with larger
$N$ more planets are liberated per relaxation time).
The values outside and inside brackets are for
the soft and hard planetary systems respectively.  In other words, we
predict for 47 Tuc, the dissolution rate is all (7 \%) of soft
(hard) planetary systems per relaxation time.
We have said that the hard time scale is 14.66
relaxation times (44.27 Gyrs, based on Table~\ref{tab:ffdat}); 
here ``soft'' refers to
planets with semi-major axes between 3 and 50 AU, while ``hard''
refers to 0.03 to 5 AU, both in logarithmically equal distribution.
\citet{Fr06} quote a lifetime of 10 Gyrs for hard planetary systems
(0.05 AU) in their statistical analysis, which is also in qualitative
agreement with earlier work of \citet{Smith01}. From these previous
papers, one would have expected that most present-day planetary
systems in globular clusters have been dissolved. 
While the given time scales are only a rough estimate, 
detailed models as presented in this paper
and also in \citet{Fr06} suggest that a significant fraction of planets with semi-major
 axes between 0.03 and 5 AU (hard in our notation, intermediate in that of \citet{Fr06})
 could have survived stellar encounters in this cluster
(see further discussions below).

The Orion nebula cluster (ONC) is an example of a young star cluster
with intermediate central density, of order $2 \cdot 10^4 M_\odot /
{\rm pc}^3$.  For this system, we use a half-mass radius of 0.8 pc and
a velocity dispersion of 2.34 km/s, to obtain the half-mass crossing
time as $3.25\cdot 10^5$ yrs. For ONC also the half-mass relaxation
time ($t_{\rm rh} = 6.5 \cdot 10^6$ yrs) and particle number $N= 2800$
are provided \citep{Hillenbrand98}. Planetary system lifetimes are then
obtained by using our measured lifetimes in units of crossing times
(see Table~\ref{tab:ffdat}).

Note that a scaling using the measured
half-mass relaxation time of the ONC would provide approximately the
same results, if the proper Coulomb scaling factor is applied to our
measured time scales (in units of relaxation time, see
Tables~\ref{tab:HMCdata},~\ref{tab:NBdata}).

For the Hyades cluster, which is a relatively low density open cluster
we use the half-mass radius $r_h = 4.5 $ pc and central velocity
dispersion of $0.3$ km/s to determine a half-mass crossing time of
approx. $1.4 \cdot 10^7$ yrs \citep{perryman98}. Again we derive
physical timescales in Table~\ref{tab:ffdat}.
Note that these estimates are subject to
many uncertainties, such as the use in our models of stars of equal mass.

\subsection{Total Cross Sections and Timescales to get Free Floaters}

\citet{Fr06} provide in their Monte Carlo simulations cross sections
for orbital changes of planetary systems
obtained from a large set of simulations of three-body
encounters. 
We want to compare their results with ours for
the case of ionization of planetary systems due to relatively close
encounters.  For that we use the usual ansatz for the planetary system
lifetime as
\begin{equation}
\tau_{\rm ff} = \frac{1}{n \sigma_{\rm ion} V }
\label{eq:cross}
\end{equation}
where $n = \rho / m$ is the average particle density, $\sigma_{\rm
ion}$ is the total cross section for dissolution of planetary systems,
and $V$ is the typical encounter relative velocity.  The relevant
total cross section for ionization in our case is taken from
equation~(\ref{eq:Delta_non-tidal}) for non-tidal impulsive encounters
as
\begin{eqnarray}
\sigma_{\rm ion} &=& \frac{8\pi Gm}{V^2} a \left[ F(\Delta_{\rm max}) 
- F(\Delta_{\rm min}) \right] \cr
F(\Delta) &=& \frac{1}{\Delta} \left( \frac{2}{3\Delta} - 1 \right)
\end{eqnarray}
where we have obtained the total cross section for ``ionization'' by
integrating equation~(\ref{eq:Delta_non-tidal}) ($\Delta < 0$) from
some suitable $\Delta_{\rm min}$ to $\Delta_{\rm max}$. With
$\Delta_{\rm max}\rightarrow -\infty$ and the assumption that
$\Delta_{\rm min} = -1$ (the minimum required relative energy change
for ionization is unity) we get $F(\Delta_{\rm max}) = 11/3$, thus
\begin{equation}
\sigma_{\rm ion} =  \frac{40\pi }{3}\frac{Gm}{V^2} a 
\end{equation}
and
\begin{equation}
\tau_{\rm ff} = \frac{3 V}{ 40\pi G \rho a }
\label{eq:tau}
\end{equation}
This result is in  agreement with \citet{Fr06} in the case of equal
stellar masses, although we
get for the case of 47 Tuc slightly larger values of the planetary
lifetimes. 
Possible reasons are that the distribution of encounter velocities in a real cluster differs
from those in three-body experiments, and that planets, which just escape from a three-body
encounter with a very small velocity at infinity could easily be recaptured by their own
host star or another star due to potential fluctuations.

Note that the lifetime of planetary systems as given in equation~(\ref{eq:tau})
does not vary with the particle number $N$, provided the planetary system's
semi-major axes remain constant relative to the $N$-body scale (system scaling
radius), consistent with our numerical results. If one is interested in
ionisation time scales relative to the relaxation time we have to look at
the quantity
\begin{equation}
\frac{1}{x_{\rm pl-ff,rh}} = \frac{\tau_{\rm ff}}{\tau_{\rm rh}} \sim 
     \frac{Gm \log\Lambda}{V^2 a}
\end{equation}
where $\log\Lambda $ is the Coulomb logarithm contained in the relaxation time. 
If we take the standard Coulomb logarithm $\log\Lambda = \log(\gamma N)$, with
$\gamma = 0.11$ (cf. \citet{GierszH94}) we get for the scaling

\begin{equation}
x_{\rm pl-ff,rh} = \frac{\tau_{\rm rh}}{\tau_{\rm ff}} \sim \frac{ V^2 a}{Gm \log(\gamma N)}
\sim \frac{ N }{\log(\gamma N)}
\end{equation}

This is indeed the empirically measured scaling behaviour, as we have commented in
the caption to Table~\ref{tab:HMCdata}. Note that in these considerations, also in the
choice of initial conditions for our runs with different particle number (HMC and
$N$-body) we have neglected the possible influence of a separate variation of the
planetary orbit scale $a$ with respect to the $N$-body system's scaling radius $R$. 
Since $\tau_{\rm ff} \sim R/a $ the effect would be easy to estimate to first order
(if only the size of planetary orbits relative to the cluster size is varied). 
If however, the scale changes are due to different cluster models (e.g.
King models of different concentration) further numerical studies would be 
required.

\subsection{Discussion}

We have provided a further step towards a self-consistent
modelling of the origin of differences in the
frequency  of planetary
systems between star clusters and the galactic field due to
encounters of planetary systems with passing stars. For two
representative star clusters with particle numbers resembling open
and globular clusters, respectively, we have
tested analytical cross
sections against two different series of
self-consistent numerical simulations under realistic conditions of a
star cluster (not merely a series of three-body scattering
experiments). 
For the illustration of our results, we have provided
liberation time scales for planets (to become a free floater) in
their respective cluster environment. In
contrast to previous work we have extended our study to planetary
systems ranging from 0.05 AU to 50 AU in semi-major axis and all
eccentricities. To ease discussion and understanding we divide our
planetary systems in two classes (hard and soft), and determine the average
time scale for their survival,
see Table~\ref{tab:ffdat}.
We find that the liberation rate of planets
per crossing time is constant, and per relaxation time scales
with the scaling factor $N/\log\Lambda $, where $\log\Lambda $ is a 
Coulomb logarithm (provided the planetary system's size remains
constant relative to the star cluster radial scale).

The results have been applied to three typical cases of clusters, the
high-density old globular cluster 47 Tuc, the intermediate density
young star forming cluster in the Orion nebula (ONC), and  the
low-density open cluster of the Hyades.  We find that even in the
dense environment of 47 Tuc about one quarter of the short-period
planets survive even after 10 Gyrs.  Our results are interesting for a
number of reasons. First of all, the diffusive nature of encounters
between planetary systems and stars has been confirmed. With the
exception of the few close encounters most intermediate and distant
encounters have no preferred sign of change for the orbital parameters
of the planet. Therefore planets can be scattered onto a more
eccentric orbit, even if they are initially very close to the host
star. Subsequent evolution of such planets by tidal effects may lead
to their inflation and tidal disruption. This
process is expected to be particularly effective in dense globular
clusters. Therefore it is not guaranteed that the fraction of
surviving systems we have found for the case of 47 Tuc would be
observable - it may just be destroyed by secondary processes such as
tidal interactions. Here we have just rescaled from our exemplary model
simulations to the particle number and mass of 47 Tuc; a detailed
case study of 47 Tuc would require a model using the half-mass
radius and concentration parameter properly tailored so as to 
reproduce the presently observed values. Also, a more detailed
case study of 47 Tuc would require as well to include a stellar mass spectrum,
stellar evolution, some binary fraction. This is subject of
future work and beyond the scope of this first exploratory study.

In contrast to the less-rich Taurus star forming region, the ONC
contains massive stars similar to the hypothetical progenitors
which generated the radioactive isotopes prior to the formation of the
solar system.  This and other, slightly less rich associations are also
likely to disperse in $\sim 100$ Myr rather than evolve into an open
cluster.  The main issue here is whether the planet-bearing stars can
preserve their companions before joining the field star population.
Our numerical results indicate that in the ONC a few times $10^8$
years is a critical threshold for the survival of wider planets and
planetary systems such as the solar system.  Thus, the disruption of
some planetary systems is expected, especially near the region where
dense protostellar cores are concentrated.  These are also the regions
where the existence of free floaters have been reported.  Our models
suggest that a significant fraction of the free-floater population may
represent the relics of disrupted planetary systems.  Our results also
indicate that the velocity dispersion of the freely floating planets
must be generally small,  as most are retained by the cluster
potential, at least for the case of soft planets (Tables
\ref{tab:HMCdata} and \ref{tab:NBdata}).

Finally, for  open clusters such as the Hyades, there is no problem in
preventing all the planetary systems (wide or compact) from destruction
before the cluster dissolves on a Gyr time scale.  The observed
dichotomy in the fraction of stars with planets between the field
stars and the Hyades requires additional explanation.  The present-day
age of the Hyades is $\tau_{Hya} \simeq$ 0.8 Gyr.  It is likely that
the cluster was somewhat denser in the past which would shorten the
planetary-system disruption time scale below that listed in Table~\ref{tab:ffdat}.
A more compact young Hyades would also reduce its $\tau_{\rm ff}$
below its lifespan.  Both eccentricity excitation and modest
migration would shake up otherwise stable planetary systems.  If
systems are driven into nearly resonant configurations, modest
eccentricity excitation can render them dynamically
unstable. We have not yet taken into account the dynamics of
multi-planetary systems in our model, but this subject will be
addressed in our future investigations.

Last, but not least, stellar encounters with planetary system can
alter the eccentricity of orbits with relatively large semi-major axes
efficiently, and also they are the only process which can create significant
amounts of inclinations out of the original orbital plane of the planetary
system (though we have not considered that issue in this paper). These
properties are potentially important for understanding the
formation of Kuiper-Belt objects.

We have neglected for the present paper the effect of stellar binaries
and multiples, as well as that of a stellar mass spectrum. In
particular the presence of a large number of binaries would be expected to have a profound
effect through the action of  binary system - planetary system encounters. The
inclusion of binary stars would certainly reduce the magnitude of
$\tau_{\rm ff}$ and the key issue is by what
extent. Also a further study of the free floaters in globular and open
clusters is interesting. Due to their very small mass they cannot
receive large amounts of energy in the mass segregation -
equipartition processes structuring the cluster during its dynamical
evolution. Therefore large numbers of free floating planets should be
retained in the cluster, a conclusion in conflict at least with
present observations.   
From the point of   view of escape, free-floating planets would 
behave very much like low-mass stars. There is much observational 
and theoretical evidence that low-mass objects escape preferentially 
from star clusters, especially in the presence of steady and time-dependent
tidal fields.  Mass segregation tends to drive such objects to large
radii, where they may be efficiently removed by tidal effects.
To study the fate of a component of very low-mass planetary objects 
during the standard evolution of star cluster is another interesting 
future project.

We conclude that, for an understanding of the diversity of planetary
systems, the fact that they originate in a dense star cluster cannot
be neglected.  Our solar system could have formed in an open cluster
or in the outskirts of an object which evolved from an ONC like star
cluster.  Other processes shaping planetary systems, such as
resonances, and interaction with gas, need to be distinguished from
the effect of diffusive encounters. 

\section*{Acknowledgments}
RS and DNCL acknowledge support from a DAAD funded personal project
partnership. RS wishes to thank Doug Lin and collegues at the UC Santa Cruz and 
the Kavli Institute for Astronomy and Astrophysics in Beijing, China for
extraordinary kind and friendly support and hospitality, and
support of the Global Networks Funds of Heidelberg University is appreciated.
The collaboration of MG and RS has been made possible by a
cooperation grant of German Science Foundation under 436 POL
113/103. This research was supported in part by NASA
(NAGS5-11779, NNG04G-191G, NNG06-GH45G), JPL(1270927),
NSF(AST-0507424, PHY99-0794) and by the Polish National Committee for
Scientific Research under grant 1 P03D 002 27.  DCH warmly thanks MG
for his hospitality during a visit to Warsaw which greatly facilitated
his contribution to this project. Simulations were performed on the 
GRACE supercomputer (Grants I/80\,041-043 of the Volkswagen
Foundation and 823.219-439/30 and /36 of the Ministry of Science, Research 
and the Arts of Baden-W"urttemberg) and on the IBM Jump
supercomputer at NIC J\"ulich. We thank the referee, John Fregeau, for a careful 
and constructive review of the paper.

\appendix

\section{Formulae of three-body scattering for planetary problems}

We present formulae for the change in binding energy and
eccentricity of a binary as a result of scattering by a distant
third body. 
This is a comprehensive collection of main results from earlier
papers (Heggie 1975, Heggie \& Rasio 1996, Roy \& Haddow 2003,
Heggie 2005), complemented by a few new results, in order to match
the requirements for comparison with our numerical simulations.

 These formulae assume that the encounter is adiabatic,
but take account of the hyperbolic geometry of the relative orbit.
In adiabatic encounters we assume that the encounter between the host and field
stars occurs on a much longer time scale than the planet's orbital
period. Modification of the perturbing potential due to the passage
of the field star is essentially adiabatic. Consequently, 
changes of the eccentricity $\delta e$ can be
computed with an orbit-averaging method, in which incremental
change of $\delta e$ per orbit can be evaluated for an
instantaneous field star position.  For $\Delta$, however, the
analytical technique is a bit more subtle (see the original papers). 

\subsection{Change of eccentricity}\label{sec:e-change}

\subsubsection{Non-circular binaries}\label{sec:non-circ}

Long ago Heggie (1975), equation~(5.66), derived a formula for the change in
eccentricity of a binary subject to a parabolic flyby of a third
body.  Corrected for an overall sign error, it is:
\begin{equation}
  \delta e =
  -\frac{15\pi}{4\sqrt{2}}\frac{m_3}{\sqrt{M_{12}M_{123}}}e\sqrt{1-e^2}
  \left(\frac{a}{r_p}\right)^{3/2}
   (\hat\ba\cdot\hat\bA\, \hat\bb\cdot\hat\bA
  + \hat\ba\cdot\hat\bB\, \hat\bb\cdot\hat\bB),
\end{equation}
where $e$ is the eccentricity, $m_3$ is the mass of the
perturber, $M_{12} = m_1+m_2$ is the total mass of the binary,
$M_{123} = m_1+m_2+m_3$ is the total mass of the system, $a$ is the
(initial) semi-major axis of the binary, $r_p$ is the distance of
closest approach between the perturber and the centre of mass of the
binary (on a Keplerian approximation), $\hat\ba$ is a unit vector
along the pericentre of the binary, $\hat\bb$ is an orthogonal unit
vector in the plane of motion of the binary (so that
$\hat\ba\wedge\hat\bb$ is directed along the angular momentum vector
of the binary), $\hat\bA$ is a unit vector along the pericentre of
the third body, and $\hat\bB$ is an orthogonal unit vector in the
plane of motion of the third body (so that $\hat\bA\wedge\hat\bB$ is
directed along the angular momentum vector of the relative motion of
the third body and the binary). Heggie \& Rasio (1996) gave the
corresponding formula for a hyperbolic flyby:
\begin{eqnarray}\label{eq:de_hyperbolic}
  \delta e &=&-{5e\over 2e^{\prime^2}} \sqrt{{m_3^2a^3(1-e^2)\over
M_{12}M_{123}\rp^3(1+e^\prime)^3}}\times\cr
&&\times
\left\{ \bahat\cdot\bAhat\, \bbhat\cdot\bAhat
\left[3e^{\prime^2}\arccos\left(-{1\over
e^\prime}\right)  + (4e^{\prime^2}-1)\sqrt{e^{\prime^2}-1}\right] \right.\cr
&&\left. + \bahat\cdot\bBhat\, \bbhat\cdot\bBhat
\left[3e^{\prime^2}\arccos\left(-{1\over
e^\prime}\right) + (2e^{\prime^2} +
1)\sqrt{e^{\prime^2}-1}\right]\right\},
\end{eqnarray}
where $e^\prime$ is the eccentricity of the third body. Expressing
the unit vectors in the equation above using orbital elements as in Roy \& Haddow
(2003), equation~(18) (angles $\omega$, $\Omega$, $i$)
we can write the result as follows:
\begin{eqnarray}
  \delta e &=&-\,{15\over 4} \Bigl({m_3^2\over M_{12}M_{123} } \Bigr)^{1/2}  \Bigl({
a \over r_p}\Bigr)^{3/2}
 {e \sqrt{1-e^2} \over (1+e')^{3/2}} \times \cr
&&\times
\left\{\sin^2 i \sin(2\Omega) \left[ \arccos(-1/e') + \sqrt{e'^2-1} \right]  +
\right. \cr
&&  \left. + {1\over 3} \left[ (1+\cos^2 i) \cos(2\omega) \sin(2\Omega)
    + 2\cos i \sin(2\omega) \cos(2\Omega)\right]
          {({e'}^2-1)^{3/2} \over {e'}^2 } \right\}\label{eq:de-noncirc}
\end{eqnarray}
where $\Omega$ is the longitude of the ascending node of the orbit
of the third body, measured in the plane of motion of the binary
from a particular origin;
$i$ is the inclination of the two orbits; and $\omega$ is the
longitude of pericentre of the third body, measured from the
ascending node. Thus we have arrived at equation~(7) of Heggie \& Rasio
(1996). In Section~2.2 and Appendix~\ref{sec:cross_sections_for_e} we
use this expression to estimate cross sections.

\subsubsection{Circular binaries}

The above formulae are all that is needed for most purposes, but for
the sake of exposition we also give here two formulae, also from
Heggie \& Rasio (1996), for the case when $e = 0$ initially.  For
hyperbolic encounters the result is:
\begin{eqnarray}
\delta e &=& 3\sqrt{2\pi} {m_3 M_{12}^{1/4}\over M_{123}^{5/4}}
    \left({\rp\over a}\right)^{3/4} {(e^\prime +1)^{3/4}\over e^{\prime^2}}\times
\cr &&
\times\exp\left[-\left({M_{12}\over M_{123}}\right)^{1/2}\left({\rp\over
a}\right)^{3/2}
{\sqrt{e^{\prime^2}-1}-\arccos(1/e^\prime)\over(e^\prime-1)^{3/2}}\right]\times
\cr &&
\times\cos^2{i\over2}\left[\cos^4{i\over2} + {4\over9}\sin^4{i\over2} +
{4\over3}\cos^2{i\over2}\sin^2{i\over2}\cos(4\omega + 2\Omega)\right]^{1/2},
\label{eq:de0circ}
\end{eqnarray}
This result simplifies a little for parabolic motion of the third
body, to
\begin{eqnarray}
\delta e &=& 3\sqrt{2\pi} {m_3 M_{12}^{1/4}\over M_{123}^{5/4}}
\left({2\rp\over a}\right)^{3/4}
\exp\left[-{2\over3}\left({2M_{12}\over
M_{123}}\right)^{1/2}\left({\rp\over a}\right)^{3/2}\right]\times
\cr &&\times\cos^2{i\over2}\left[\cos^4{i\over2}
 + {4\over9}\sin^4{i\over2} +
{4\over3}\cos^2{i\over2}\sin^2{i\over2}\cos(4\omega +
2\Omega)\right]^{1/2}.\label{eq:de-circ-parabolic}
\end{eqnarray}

\subsection{Change of binding energy}

\subsubsection{Non-circular binaries}

Roy \& Haddow (2003) give the following expression for the change in
energy of the binary in the case of a parabolic flyby:
 \begin{eqnarray}
   \delta\varepsilon &=& -
   \frac{Gm_1m_2m_3}{M_{12}r_p^3}\frac{\sqrt{\pi}}{120}K^{5/2}\exp(-2K/3) \times\cr
&& \times \left\{ 60a^2e_1
     \left(\sin (nt_0)(\bahat.\bBhat)^2
   +2\cos (nt_0)\bahat\cdot \bAhat\, \bahat\cdot \bBhat
   - \sin (nt_0)(\bahat\cdot \bAhat)^2\right) + \right.\cr
&&\left.+120abe_4 (-\sin (nt_0)\bbhat\cdot \bBhat\, \bahat\cdot \bAhat
             -\sin (nt_0)\bahat\cdot \bBhat\, \bbhat\cdot \bAhat + \right.\cr
         &&  \left. +\cos (nt_0)\bahat\cdot \bBhat\, \bbhat\cdot \bBhat
             -\cos (nt_0)\bahat\cdot \bAhat\, \bbhat\cdot \bAhat) + \right.\cr
&& \left. +60b^2e_2
   \left(\sin (nt_0)(\bbhat\cdot \bAhat)^2
       -2\cos (nt_0)\bbhat\cdot \bBhat\, \bbhat\cdot \bAhat
       - \sin (nt_0)(\bbhat\cdot \bBhat)^2\right)  \right\},
\label{eq:deps-p}
 \end{eqnarray}
where $K = \displaystyle{\sqrt{\frac{2M_{12}r_p^3}{M_{123}a^3}}}$,
$b = a\sqrt{1-e^2}$ is the semi-minor axis of the binary, $n =
\displaystyle{\sqrt{\frac{GM_{12}}{a^3}}}$ is the mean motion of the
binary, $t_0$ is the time of pericentric passage of the binary (on a
time scale in which closest approach of the third body occurs at
$t=0$), and the coefficients are given by:
\begin{eqnarray}
  e_1 &=& J_{-1}(e) -2eJ_0(e) + 2eJ_2(e) - J_3(e)\cr
  e_2 &=& J_{-1}(e) - J_3(e)\cr
  e_3 &=& eJ_{-1}(e) -2J_0(e) + 2J_2(e) - eJ_3(e)\cr
\mbox{}  e_4 &=& J_{-1}(e) -eJ_0(e) - eJ_2(e) + J_3(e);
\end{eqnarray}
here $J_n$ is the Bessel function of the first kind of order $n$.

By combining the mathematical procedures in the two papers Heggie \&
Rasio (1996) and Roy \& Haddow (2003) it is not hard to show that
the corresponding result for a hyperbolic encounter of eccentricity
$e^\prime$ gives the following expression:
 \begin{eqnarray}
   \delta\varepsilon &=& -
   \frac{Gm_1m_2m_3}{M_{12}a^3}\frac{\sqrt{2\pi}}{120}\left(
\frac{M_{12}}{M_{123}}\right)^{5/4}\left(
\frac{r_p}{a}\right)^{3/4}\frac{(e^\prime+1)^{3/4}}{e^{\prime2}}\cr
&&\exp\left(-\frac{n}{n^\prime}\left(\sqrt{e^{\prime2}-1}
   - \arccos\frac{1}{e^\prime}\right)\right)\left\{60a^2e_1\left(\sin
   (nt_0)(\bahat\cdot \bBhat)^2 +\right.\right.\cr
&&\left.+2\cos (nt_0)\bahat\cdot \bAhat\, \bahat\cdot \bBhat - \sin
   (nt_0)(\bahat\cdot \bAhat)^2\right) + \cr
&&+120abe_4(-\sin (nt_0)\bbhat\cdot \bBhat\, \bahat\cdot \bAhat - \sin
   (nt_0)\bahat\cdot \bBhat\, \bbhat\cdot \bAhat +\cr
&&+\cos (nt_0)\bahat\cdot \bBhat\, \bbhat\cdot \bBhat -\cr &&-\cos
(nt_0)\bahat\cdot \bAhat\, \bbhat\cdot \bAhat) + 60b^2e_2\left(\sin
   (nt_0)(\bbhat\cdot \bAhat)^2 -\right.\cr
&&\left.{\left.-2\cos (nt_0)\bbhat\cdot \bBhat\, \bbhat\cdot \bAhat -
\sin (nt_0)(\bbhat\cdot \bBhat)^2\right)}\right\};
 \end{eqnarray}
here $n^\prime =
\displaystyle{\sqrt{\frac{GM_{123}}{a^{\prime3}}}}$, where
$a^\prime$ is the semi-major axis of the hyperbolic motion, and so
$r_p = a^\prime(e^\prime-1)$. Now expressing again the result by
using orbital elements we get:
 \begin{eqnarray}
   \delta\varepsilon &=& -
   \frac{Gm_1m_2m_3}{M_{12}r_p^3}\frac{\sqrt{\pi}}{8}
\frac{(e^\prime+1)^{3/4}}{2^{3/4} e^{\prime2}} K^{5/2}
     \exp\left(-{K\over\sqrt{2}}\left(\frac{\sqrt{e^{\prime2}-1}
   - \arccos(1/e^\prime)}{(e^\prime-1)^{3/2}}\right)\right) \times \cr
&& \times \left\{ e_1 a^2 \left[ \sin(2\omega +nt_0)(\cos(2i)-1) - \sin(2\omega
+nt_0)\cos(2\Omega)\cos(2i) - \right.\right. \cr
&& \left.\left. -3 \sin(2\omega +nt_0)\cos(2\Omega) - 4 \cos(2\omega +nt_0)
\sin(2\Omega) \cos i \right] + \right .\cr
&& \left. +e_2b^2 \left[ \sin(2\omega +nt_0)(1-\cos(2i)) - \sin(2\omega
+nt_0)\cos(2\Omega)\cos(2i) - \right.\right. \cr
&& \left.\left. - 3\sin(2\omega +nt_0)\cos(2\Omega) -4\cos(2\omega +nt_0) 
\sin(2\Omega) \cos i \right] + \right .\cr
&& \left. + e_4ab \left[ -2\cos(2\omega +nt_0)\sin(2\Omega)\cos(2i) -6 \cos(2\omega
+nt_0)\sin(2\Omega) - \right.\right. \cr
&& \left.\left. - 8 \sin(2\omega +nt_0)\cos(2\Omega) \cos i \right] \right\}
 \label{eq:deps}
  \end{eqnarray}
where $\Omega$ is measured from $\bahat$.
From here the interested reader should read off the definition of
the functions $F$, $f_1$, and $f_2$ used in equations~(\ref{eq:Delta}) and
(\ref{eq:f1f2}).

\subsubsection{Circular binaries, parabolic case}

Roy and Haddow (2003) also provided an expression for the energy change
of a circular binary in the case of parabolic flyby. This formula,
expressed in terms of  the orbital elements, is as follows:
 \begin{eqnarray}
   \delta\varepsilon &=& -
   \frac{Gm_1m_2m_3}{M_{12}}
\frac{\sqrt{\pi}a^3}{8 r_p^4}K^{7/2}\exp\left(-{2\over3} K\right)(\mu_2-\mu_1)
 (1 + \cos i)\sin^2 i  \quad \times \cr
&& \times [(\cos^3 \omega - 3\sin^2 \omega \cos \omega) \sin(nt_0) +
(3\cos^2 \omega \sin \omega - \sin^3 \omega)\cos(nt_0)]
 \label{eq:cdeps}
  \end{eqnarray}
where $\mu_i = m_i/M_{12} \quad (i = 1,2)$ are the reduced masses.

\section{Cross sections for the change of
eccentricity}\label{sec:cross_sections_for_e}

\subsection{Impulsive tidal encounters}\label{sec:imp_tid}

Here our starting point is an expression for the total change, due to
the passing star, in the specific angular momentum $J$ of the planet
relative to its sun.  Following \citet{H75}, where (in Section~4.2) an
analogous calculation of the energy change is given, we obtain
\begin{equation}
  \delta \bJ = \frac{2Gm}{V}\br\times\left(\frac{\bp - \br_\perp}{\vert\bp -
  \br_\perp\vert^2} - \frac{\bp}{\vert\bp\vert^2}\right).
\end{equation}
Here $\bp$ is a vector from the sun to the closest point of the
(rectilinear) path of the passing star, and $\br_\perp$ is the
component of the position vector of the planet, at this time, which is
orthogonal to the path of the passing star.  As usual we have
specialised to the case of equal stellar mass and negligible planetary
mass.

In the tidal regime we expand in powers of $\br_\perp$, obtaining to
lowest order 
\begin{equation}\label{eq:dbJ}
  \delta \bJ = \frac{2Gm}{Vp^2}\br\times\left(-\br_\perp + 2(\br_\perp
\bphat)\bphat\right),
\end{equation}
where $\bphat$ is the unit vector parallel to $\bp$.  Hence
\begin{equation}
  \bJ.\delta\bJ = - \frac{2Gm}{Vp^2}\mybv.\br\times(\br\times\left(-\br_\perp + 
(\br_\perp.\bphat)\bphat\right)).
\end{equation}

The
corresponding result for the change in energy, $\delta\varepsilon$, is
similar, except that the factor $\br\times$ in equation~(\ref{eq:dbJ}) is replaced by
$\mybv\cdot~$,
where $\mybv$ is the velocity of the planet (relative to its sun) at the
time of closest approach of the passing star.  By equation~(\ref{eq:dJ}) the
change in $e^2$ can now be written as 
\begin{equation}
  D = \frac{4r_\perp}{GmVp^2}\bA_\perp.(-\hat{\br_\perp} + 2\hat{\br_\perp}
\bphat\bphat)
\end{equation}
where $\bA = J^2\mybv + 2\varepsilon\bJ\times\br$.  Now
$\bA_\perp,\br_\perp$ and $\bphat$ are vectors in the same plane, and
so $D = -\displaystyle{\frac{4r_\perp A_\perp}{GmVp^2}\cos(\psi +
  2\theta)}$, where $\theta,\psi$ are the angles in this plane from
$\hat{\br_\perp}$ to $\bphat, \bA_\perp$, respectively.  Thus the
differential cross section, before averaging all the binary
parameters, is
\begin{equation}
  \frac{d\sigma}{dD} = \int\delta\left(D + 
\frac{4r_\perp A_\perp}{GmVp^2}\cos(\psi + 2\theta)\right) p dp d\theta.
\end{equation}
Hence 
\begin{eqnarray}
  \frac{d\sigma}{dD} &=& \int\frac{p
  dp}{\sqrt{\displaystyle{\left(\frac{4r_\perp A_\perp}{GmVp^2}\right)^2 - D^2}}}\\
&=& \frac{2r_\perp A_\perp}{GmVD^2}.
\end{eqnarray}

Next we note that $r_\perp$ scales as $a$, the semi-major axis, and
$A$ scales with $(Gm)^{3/2}a^{1/2}$.  Hence we can write 
\begin{equation}
  \frac{d\sigma}{dD} = \frac{2(Gm)^{1/2}a^{3/2}}{VD^2}r_{\perp1} A_{\perp1},
\end{equation}
the extra subscripts denoting these normalisations.  

This is as far as we have been able to take the analytic evaluation of
this cross section.  Resorting to Monte Carlo integration to average
over the normalised binary parameters, we found that 
\begin{equation}
  \frac{d\sigma}{dD} = \frac{2(Gm)^{1/2}a^{3/2}C_1}{VD^2},
\end{equation}
where $C_1 = 0.883$ to three significant figures.

\subsection{Extremely hyperbolic adiabatic encounters}\label{sec:D_hyperbolic}

We begin with equation~(\ref{eq:de_hyperbolic}), let $m_1 = m_3 = m$ and
$m_2=0$, and take the asymptotic form for $e'\to\infty$.  In this
limit $r_p\simeq p$, the impact parameter, and we evaluate the
limiting expression for $e'$ using equation~(\ref{eq:eprime}).  The result
is 
\begin{equation}
  \delta e =
  -\frac{5e\sqrt{Gm(1-e^2)}a^{3/2}}{Vp^2}(2\bahat.\bAhat\bbhat.\bAhat + \bahat
\bBhat\bbhat.\bBhat),
\end{equation}
where $\bahat, \bbhat$ are certain orthogonal unit vectors in the plane
of motion of the planet, and $\bAhat, \bBhat$ are similarly defined in
the plane of motion of the passing star (Appendix~\ref{sec:non-circ}).  Hence the
total cross
section for $D$ is, for $D_0 >0$, 
\begin{equation}
  \sigma(D>D_0) = \pi p^2 =   - 
\frac{10\pi e^2\sqrt{Gm(1-e^2)}a^{3/2}}{DV}(2\bahat.\bAhat\bbhat.\bAhat + \bahat
\bBhat\bbhat.\bBhat),
\end{equation}
provided that the last factor is negative.

Now we average over the orientation of the planetary orbit.  To do
this we think of $\bAhat$ and $\bBhat$ as providing the first two
vectors of a basis, and the expression to be averaged is therefore 
$X = 2\ahat_1\bhat_1 + \ahat_2\bhat_2$, where this is negative.  With the
usual spherical polar coordinates we take
\begin{equation}
  \bahat = (\sin\theta\cos\phi,\sin\theta\sin\phi,\cos\theta),
\end{equation}
and then $\bbhat$, which is an arbitrary unit vector orthogonal to
$\bahat$, may be written as
\begin{equation}
  \bbhat = (\cos\theta\cos\phi\cos\psi - \sin\phi\sin\psi,
\cos\theta\sin\phi\cos\psi + \cos\phi\sin\psi,-\sin\theta\cos\psi).
\end{equation}
Therefore $X = \alpha\cos\psi + \beta\sin\psi$, where
\begin{eqnarray}
  \alpha &=& \sin\theta\cos\theta(2\cos^2\phi + \sin^2\phi),\\
  \beta &=& - \cos\phi\sin\phi\sin\theta.
\end{eqnarray}
Averaging with respect to $\psi$ for $X<0$ gives 
\begin{equation}
\langle X\rangle =
- \frac{2}{\pi}\sqrt{\alpha^2+\beta^2}.  
\end{equation}
It appears that averaging over $\theta$ and $\phi$ has to be done by
numerical quadrature, and yields 
\begin{equation}
\langle X\rangle =  -\frac{2C_2}{\pi}
\end{equation}
where $C_2 = 0.5932$ approximately.  

Finally we average over a thermal distribution of $e$, and
differentiate with respect to $D$ to obtain the differential cross
section
\begin{equation}
  \frac{d\sigma}{dD} = \frac{16C_2}{3}\frac{\sqrt{Gma^3}}{VD^2}.
\end{equation}

\subsection{Near-parabolic encounters}\label{sec:D_parabolic}

We begin with equation~(19) in \citet{HR96}.  Specialising to the case of
two equal masses $m$ and one vanishing mass, this becomes
\begin{equation}
  \sigma(\delta e>\delta e_0) =
  \frac{9\sqrt{3}}{14}\left(
\frac{15\pi}{2}\right)^{2/3}\left[\Gamma\left(
\frac{2}{3}\right)\Gamma\left(
\frac{5}{6}\right)\right]^2\frac{Gma}{V^2}e^{2/3}(1-e^2)^{1/3}(\delta e_0)^{-2/3}.
\end{equation}
Eliminating $\delta e$ in favour of $D = \delta(e^2)\simeq 2e\delta
e$, and then differentiating with respect to $D$, we arrive at the
differential cross section
\begin{equation}
  \frac{d\sigma}{dD} = \frac{3\sqrt{3}}{7\pi}(15\pi)^{2/3}
\left[\Gamma\left(\frac{2}{3}\right)\Gamma
\left(\frac{5}{6}\right)\right]^2
\frac{Gma}{V^2}e^{4/3}(1-e^2)^{1/3}(D)^{-5/3}.
\end{equation}
Finally, averaging over a thermal distribution of eccentricities,
i.e. $f(e) = 2e$, gives
\begin{equation}
  \frac{d\sigma}{dD} = \frac{2}{21}(15\pi)^{2/3}
\left[\Gamma\left(\frac{2}{3}\right)\Gamma
\left(\frac{5}{6}\right)\right]^2\frac{Gma}{V^2}(D)^{-5/3}.
\end{equation}

\section{Cross Sections for the Change in Energy}\label{sec:Delta_cross_sections}

Here we consider only adiabatic encounters, and turn first to the parabolic
case. Equation~(\ref{eq:deps-p}) shows that 
\begin{equation}
  \Delta = g\exp\left(-\frac{2K}{3}\right),
\end{equation}
where $K = (r_p/a)^{3/2}$ for the present assumptions about the masses,
  and $g$ is a dimensionless coefficient which depends on K and on the
  geometry of the encounter.  Remarkably, we do not need to evaluate
  it in order to obtain the cross section to lowest order.  Using
  equation~(\ref{eq:deps-p}) in the parabolic limit, we deduce that 
  \begin{equation}\label{eq:sigma_working}
    \sigma = \frac{\pi p^2}{2} = \frac{2\pi Gma}{V^2}\left(-
\frac{3}{2}\ln\frac{\Delta}{g}\right)^{2/3},
  \end{equation}
where the factor of $1/2$ in the second expression is needed because
the geometric factor $g$ has the correct sign in only half of all
encounters.  Differentiating with respect to $\Delta$, we can ignore
the derivative of $g$; $g$ depends on powers of $r_p$ or $p$ and hence
on powers of $\ln\Delta$, as we can see from
equation~(\ref{eq:sigma_working}), and so it varies much more slowly than
$\Delta$.  Similarly we can neglect $\ln\vert g\vert$ by comparison
with $\vert\Delta\vert$ for sufficiently weak encounters.  Therefore
\begin{equation}
  \frac{d\sigma}{d\Delta} = \frac{2\pi
  Gma}{V^2}\frac{1}{\vert\ln\vert\Delta\vert\vert^{1/3}\vert\Delta\vert} 
\end{equation}

The extremely hyperbolic case follows a similar line of argument.  As
stated in equation~(\ref{eq:Delta_hyp}) we have 
\begin{equation}
  \Delta = g\exp\left(-\frac{2K}{3\sqrt{e'}}\right).
\end{equation}
Now we use equations~(\ref{eq:eprime}) and (\ref{eq:prp}) to express
everything in terms of $p$, and find that
\begin{equation}
  \sigma = \frac{\pi p^2}{2} = \frac{\pi V^2a^3}{2Gm}(\ln(\Delta/g))^2.
\end{equation}
Differentiation with respect to $\Delta$ with the same approximations
leads to the result
\begin{equation}
  \frac{d\sigma}{d\Delta} = \frac{\pi
  a^3V^2}{Gm}\frac{\vert\ln\vert\Delta\vert\vert}{\vert\Delta\vert}. 
\end{equation}

\end{document}